\documentclass[11pt,a4paper,final]{article}
\usepackage{jheppub}
\usepackage[latin1]{inputenc}

\usepackage{bbm,bm}
\usepackage{graphicx}
\usepackage{wrapfig}
\usepackage{amssymb}
\usepackage{mathrsfs}
\usepackage{amsmath}
\usepackage{comment}
\usepackage{amsbsy}
\usepackage{mathrsfs}
\usepackage{amsfonts}
\usepackage{epsf,color,colordvi,pifont}
\usepackage{lineno}
\usepackage{nicefrac}
\usepackage[notref,notcite]{showkeys} 

\DeclareFontFamily{OT1}{pzc}{}
\DeclareFontShape{OT1}{pzc}{m}{it}{<-> s * [1.10] pzcmi7t}{}
\DeclareMathAlphabet{\mathpzc}{OT1}{pzc}{m}{it}
\DeclareMathOperator\arctanh{arctanh}

\usepackage{url}

\title{Light dark matter candidates in intense laser pulses I: paraphotons and fermionic minicharged particles}

\author{S.  Villalba-Ch\'avez}

\author{and C.  M\"{u}ller}

\affiliation{Institut f\"{u}r Theoretische Physik I, Heinrich Heine Universit\"{a}t D\"{u}sseldorf\\ Universit\"{a}tsstr. 1, 40225 D\"{u}sseldorf, Germany}

\emailAdd{selym@tp1.uni-duesseldorf.de}
\emailAdd{c.mueller@tp1.uni-duesseldorf.de}

\abstract{Polarimetric  experiments driven by the strong field  of a circularly polarized  laser wave  can become a powerful  tool  to  limit  
the parameter space of  not yet detected  hidden-photons  and  minicharged particles associated with  extra  $\rm U(1)$ gauge  symmetries.  We 
show how the absorption and dispersion of  probe electromagnetic waves in the vacuum polarized  by  such a  background   are modified  due to 
the coupling between the  visible $\rm U(1)$-gauge  sector  and  these hypothetical degrees of freedom. The results  of this  analysis  reveal  
that the  regime close to the two-photon reaction threshold  can be a sensititive  probe of these hidden particles. Parameters of modern laser systems  
are used to estimate the constraints on the corresponding coupling constants  in regions where experiments driven by dipole magnets are less 
constricted. The role played by a paraphoton field is analyzed via a comparison with  a  model  in which  the existence of  minicharges is  
assumed only.  For both scenarios is found that  the most stringent  exclusion limit occurs  at the  lowest threshold mass; this one being  
determined by  a certain combination of the field frequencies and  dictated by  energy momentum balance of the photo-production of a pair of 
minicharged particles. The dependencies of the observables on the laser  attributes as well as  on the unknown  particle parameters are  also analyzed.
}


\begin{document}
\maketitle
\flushbottom

\section{Introduction}

String theory encompasses at  present some of the most promising candidates for a unified description of the fundamental 
forces in nature. Four-dimensional  remnants resulting from the compactification of extra dimensions, provide a variety 
of Standard Model (SM) extensions which very often contain--in addition to the  $\rm SU(3)\times SU(2)\times U(1)$  gauge 
group--a hitherto unobserved very weakly interacting sector involving further gauge invariances \cite{Witten:1984dg,Lebedev:2008un,Lebedev:2009ag,Goodsell:2009xc}. 
The derivation of effective theories with such symmetry properties opens a portal for the  insertion  of hidden degrees of 
freedom whose  realization in nature might be linked to the abundant dark  matter and dark energy  of our universe \cite{Jaeckel:2010ni,Ringwald:2012hr,Hewett:2012ns,Essig:2013lka}. 
Determining the extent to which these outcomes adjust to a realistic description of the latter subject is a fundamental task 
in particle physics. Mainly, because it could not only reveal why  other puzzles in the SM lack  a satisfactory theoretical  
explanation but could also validate  the  building blocks on which  it relies. Notably, with respect to  the charge quantization,   
as of today it is still not clear whether  or not it represents  a fundamental principle. Indeed, the conclusion resulting from some effective scenarios 
is that this might not be the case once the hidden sector includes  an extra  $\rm U(1)$ symmetry and  the respective   paraphoton 
\cite{Okun:1982xi,Langacker:2008yv,Ahlers:2007rd,Ahlers:2007qf} or hidden-photon field--minimally coupled to very light particles 
under the same $\rm U(1)$ group--is kinetimatically mixed with the visible electromagnetic sector. It is precisely the diagonalization 
of this term  what allows us to predict carriers with unquantized  electric charges. In  this context,  a  presumably feeble 
interaction  leads  to introduce Mini-Charged Particles (MCPs) \cite{Holdom:1985ag,Dobrescu:2006au,Gies:2006ca,Ahlers:2006iz}, a 
concept  which  often arises in many branches beyond the SM  \cite{Batell:2005wa,Bruemmer:2009ky,Dudas:2012pb}.

Quantum Electrodynamics (QED) ammended with the  incorporation  of the hidden-photon field and very light MCPs, acquires a source of quantum 
fluctuations which could induce nonlinear interactions in the electromagnetic field as those mediated by virtual electron-positron 
pairs \cite{Dobrich:2009kd,Gies:2007ua}. Because of this fact, its phenomenology can be  modified and  experiments in strong-external 
fields--searching for elusive phenomena such as birefringence and  dichroism of the vacuum \cite{Dittrich,Hattori,VillalbaChavez:2012ea}, 
discrepancies in the Coulomb law \cite{Jaeckel:2009dh,Jaeckel:2010xx} or the generation of visible  photons from  these hypothetical 
degrees of freedom  \cite{Redondo:2010dp,Arias:2010bh}--can become illuminating tools for testing their occurence  in nature. Inspired by this fact,  several  laboratory-based  experiments searching  for signatures of MCPs, paraphotons and  axionlike particles have 
been carried  out. Indeed,  on the basis of the mentioned optical properties of the polarized vacuum, collaborations such as BFRT 
\cite{Cameron:1993mr}, PVLAS \cite{Zavattini:2007ee,DellaValle:2013xs}, BMV \cite{BMVreport},  and Q $\&$ A \cite{Chen:2006cd} have 
performed high-precision polarimetric measurements  on  a low-energy photon beam which traverses a magnetic field  region. Likewise, based on the 
idea of photon regeneration,  many ``Light Shining Through a Wall'' experiments  have been put forward \cite{Ehret:2010mh,Ehret:2009sq,Chou:2007zzc,Steffen:2009sc,Afanasev:2008jt,Afanasev:2006cv,Pugnat:2007nu,Robilliard:2007bq,Fouche:2008jk},  
but in none of these setups  a  weakly interacting sub-eV particle has been detected so far.  Instead, the range of the  unknown 
particle masses and coupling constants has been  constricted. 

Currently,  the record for  the most stringent bounds on MCPs parameters [e. g., relative charge parameter $\epsilon\lesssim 10^{-14}$ for masses below a few  $\rm keV$]  
result from  arguments related to  stellar cooling   \cite{davidson} which are  not observed in Horizontal Branch stars. This exclusion 
limit is, nonetheless,  somewhat  arguable  since the inclusion of  macroscopic  parameters such as the density and the temperature  of 
the star might  attenuate it  significantly 
\cite{Masso:2006gc,evading}. Such  observations  motivate  the interest in  laboratory searches, ideally, with enough sensitivity as to 
compete with  the  astrophysical bounds. However, due to technical limitations, the laboratory tests via polarimetry and ``Light Shining 
Through a Wall''  do not yet reach this goal.  The main difficulty  stems from a presumably  feeble coupling  between these particles  
and the magnetic field $\vert\pmb{B}\vert \lesssim 10^5\ \mathrm{G}$,  which can be  effectively extended up to distances on  the order of $L\sim 1\ \mathrm{km}$ 
by using high-finesse interferometry.   So, to make manifest the existence of such degrees of freedom,  a significant improvement in the 
field strength as well as in the mentioned techniques are required. Meanwhile, the  upper limits derived  from the 
outcomes of the cited experimental collaborations  remain many orders of magnitude  bigger than the astrophysical one. While  a new 
generation of ``Light  Shining  Through a Wall''  experiments might overcome  this obstacle  \cite{Sikivie:2007qm,Mueller:2009wt,Bahre:2013ywa}, 
there is a real  demand for  new  theoretical efforts  toward the search of complementary scenarios  where potential  improvements in 
the bounds  of  parameters of these dark matter candidates  can be achieved \cite{Gies:2006hv,Dobrich:2012sw,Dobrich:2012jd}.  

Despite the disadvantage introduced by their limited  temporal and spatial extensions,   the prospect of finding  stringent  limits   
on the  weakly interacting sub-eV particles'  attributes  by using high-intensity laser  fields is   becoming  a subject of  interest 
\cite{Gies:2008wv,Dobrich:2010hi,Dobrich:2010ie,Tommasini:2009nh,Villalba-Chavez:2013bda,Villalba-Chavez:2013gma,Villalba-Chavez:2013goa}.  
Firstly,  because  the  field strengths attained  from these powerful sources are  much higher than the static ones  frequently  
used in  experiments driven by dipole magnets $\vert\pmb{B}\vert\sim\mathpzc{o}[10^{4}-10^{5}]\ \rm G$;  and secondly,  because  
the oscillating nature of laser fields  introduces--apart from the profile of the  wave amplitude and its polarization--the field  
frequency as  an additional characteristic. As a consequence, the processes  occuring inside  these kind of  backgrounds  are 
typified by thresholds  and resonances related to the  exchange of energy  between the quantum fields and the classical laser wave.  In connection,  a  
diversity  of  processes involving a frequency shift  of probe beams are predicted to occur   leading  to   introduce   novel  
detection techniques  such as the  Raman spectroscopy \cite{Dobrich:2010hi,Villalba-Chavez:2013gma}.  Indeed, some  phenomenological studies in this direction  are pointing 
out that the  nonobservation of these inelastic scattering waves  could  allow us  to  limit the parameter space of  axionlike 
particles and  MCPs in regions for which  the current laboratory-based experiments  establish less stringent  constraints. 

Clearly, investigations of this nature are also stimulated by ongoing projects such as the Extreme Light Infrastructure (ELI) 
\cite{ELI} and the Exawatt Center for Extreme Light Studies (XCELS) \cite{xcels}. The--so far inaccessible--field strengths to 
be attained in these high-intensity laser facilities  $\vert\pmb{B}\vert\sim\mathpzc{o}[10^{11}-10^{12}]\ \rm G$ offer a genuine opportunity for studying  
the low energy sector of particle physics as well as for observing--among other hitherto  undetected nonlinear QED phenomena 
\cite{Heinzl,DiPiazza:2007yx,nonlin1,Hatsagortsyan:2011bp,Di_Piazza_2012,King:2014vha}--the spontaneous production 
of electron-positron pairs \cite{Schwinger:1951nm,Hebenstreit:2009km,Akal:2014eua}. While this constitutes a 
very strong motivation, the first estimates of the upper  bounds  resulting  from  operating facilities such as  the  Petawatt High-Energy Laser for heavy Ion eXperiments (PHELIX) \cite{PHELIX}  
and  the Laboratoire pour l'Utilisation des Lasers Intenses  (LULI) \cite{LULI} might turn out to be competitive  in the search for MCPs and even more  
promising than those derived from ELI and XCELS parameters.  This has  already been predicted   theoretically  for axionlike particles  
\cite{Villalba-Chavez:2013goa}. The main reason  behind this finding  lies   in the  fact that these contemporary  systems--although operating at moderate 
intensity $I\sim \mathpzc{o}[10^{14}-10^{16}]\ \rm W/cm^2$--deliver  
relatively  long pulses $\tau\sim \mathpzc{o}[\rm ns]$. For axionlike particles the relevant  combination $I\tau^2\sim\mathpzc{o}[10^{14}-10^{16}]\ \rm W\ ns^2/cm^2$  
increases the sensitivity in  polarimetric experiments when  compared with the outcomes resulting from  the ELI and XCELS parameters  
at which the  temporal lengths  $\tau\sim\mathpzc{o}[\mathrm{fs}]$  significantly compensate the beneficial aspects introduced by  
the expected  high intensities $I\sim\mathpzc{o}[10^{25}-10^{26}]\ \rm W/cm^2$ [$I\tau^2\sim\mathpzc{o}[10^{13}-10^{14}]\ \rm W\ ns^2/cm^2$].

Against this background, the present   work  aims   to  provide a first estimate on the exclusion limits for MCPs  and massless 
paraphotons, resulting from plausible polarimetric setups utilizing the  field of a circularly polarized  laser wave of long 
temporal length. To this end, we first determine how the vacuum  refraction indices  and  the photon absorption coefficients  that follow from the 
vacuum polarization tensor  mediated by fermionic MCP pairs are modified   by a hidden-photon field. We find that the birefringence and dichroism  of the vacuum are quite pronounced in a vicinity of 
the first  photo-production threshold.  Our analysis reveals that--at moderate laser intensities--high-precision polarimetric experiments might  be 
sensitive probes  of  these hidden degrees of freedom.  Parameters of  the   aforementioned laser facilities  are   used for 
establishing  upper bounds on the respective parameter  spaces. The role played by a paraphoton field is analyzed via a comparison with  a  model  in which  the existence of MCPs only is  
  assumed.  For both scenarios is found that  the most stringent  exclusion limit occurs  at the  lowest threshold mass; this one being  
determined by  a certain combination of the laser frequencies and  dictated by  energy momentum balance of the photo-production of a pair of 
minicharged particles. An analysis of the signals dependence on the laser attributes as well as the on the unknown  parameters is  also included.   

\section{Loop-induced photon-paraphoton oscillations}
\subsection{Kinetic mixing and  effective action}
So far  there are no experimental evidences which indicate  a  violation  of  any  fundamental principle of QED. Hence,  we  will  consider the most  simple  renormalizable  Lagrangian 
density that  includes  both an  electromagnetic field   $a_\mu(x)$  and a  hidden vector field  $w_\mu(x)$  but   preserves  the Lorentz  invariance, the spatial parity, the  
temporal  reversibility  and  the   charge conjugation symmetry.  Furthermore, we wish  to  guarantee the gauge invariance of  the involved  fields and avoid the  proliferation of an 
additional charge  labeling   the  elementary standard-model particles. In order to satisfy these two conditions, we deal with a  theory  invariant under a $\rm U(1)\times U(1)-$gauge   
symmetry group and assume that the interaction between  both Abelian  sectors occurs through a  kinetic-mixing term characterized by a completely  arbitrary dimensionless parameter $\chi_0$.  With 
these details in mind, the  Lagrangian density turns out to be \cite{Jaeckel:2010ni,Ringwald:2012hr,Hewett:2012ns,Essig:2013lka}\footnote{From now  on ``natural'' and Gaussian  units with  $c=\hbar=4\pi\epsilon_0=1$ are used.} 
\begin{equation}\label{startingLagrangian}
 \mathpzc{L}=-\frac{1}{16\pi}f_{\mu\nu}f^{\mu\nu}-\frac{1}{16\pi}h_{\mu\nu}h^{\mu\nu}-\frac{1}{8\pi}\chi_0f_{\mu\nu}h^{\mu\nu}-e_hj_{h}^\mu w_\mu,
\end{equation}where  $f_{\mu\nu}=\partial_\mu a_\nu-\partial_\nu a_\mu$  and  $h_{\mu\nu}=\partial_\mu w_\nu-\partial_\nu w_\mu$  refer  to the corresponding field   tensors;  $j_h^\mu$   and  
$e_h$ is the  hidden current and gauge  coupling respectively  associated  with  hypothetical    particles   charged  under the extra  $\rm U(1)$ symmetry.  The explicitly expression of $j_h^\mu$ 
depends on the nature of the hidden matter sector. Hereafter we suppose that it is determined by Dirac fermions. The  kinetic mixing in Eq.~(\ref{startingLagrangian}) can be diagonalized by 
changing the hidden gauge field to another   basis  $w_\mu\to w_\mu-\chi_0a_\mu$. After having used  
the sequence of redefinitions  $a_\mu\to (1-\chi_0^2)^{-\nicefrac{1}{2}}a_\mu$  and  $\chi_0\to\chi(1-\chi_0^2)^{\nicefrac{1}{2}}$ we 
end up with
\begin{equation}\label{startingLagrangianII}
 \mathpzc{L}=-\frac{1}{16\pi}f_{\mu\nu}f^{\mu\nu}-\frac{1}{16\pi}h_{\mu\nu}h^{\mu\nu}-e_hj_h^\mu w_\mu +\chi e_hj_h^\mu a_\mu.
\end{equation}Manifestly, the last term defines  an interacting vertex which links  the hidden  matter  sector and the electromagnetic field. As a consequence,   the 
hypothetical particles acquire  an  electric charge  under the visible $\rm U(1)$-gauge field given by 
\begin{equation}
\mathpzc{q}_\epsilon\equiv\epsilon e=- \chi e_h.\label{chargesrelation}
\end{equation} Since $\chi$ is an arbitrary number, 
the parameter $\epsilon$--which acccounts for  the potentially small coupling strengh in units of the electron charge $e$--is not necessarily an integer number.  For small values of
$\chi\ll1$, one finds that $\vert\epsilon\vert\ll1$ due to which  the weakly interacting  charge  carriers are called  MCPs. 

It is opportune to emphasize that  the hidden-photons can, in general, acquire a mass term $\sim m_{\gamma^\prime}^2w_\mu w^\mu$ through the Higgs mechanism   leading to   a 
break down of the initial hidden $\rm U(1)-$symmetry \cite{Jaeckel:2010ni,Ringwald:2012hr,Hewett:2012ns,Essig:2013lka}. With the change of basis that brings the kinetic-mixing 
term  to a diagonal form,  the visible photons  become  massive particles   $\sim \chi^2 m_{\gamma^\prime}^2 a_\mu a^\mu$  
and the resulting  Lagrangian density $\mathpzc{L}$  is no longer gauge invariant. The  aforementioned transformation creates, in addition,  a massive-mixing term  
$\sim -\chi m_{\gamma^\prime}^2  a_\mu w^\mu$ which can   drive  the photon-paraphoton oscillations. However, we are  motivated to investigate such a phenomenon  mediated by  a loop  diagram 
of MCPs  rather than the  previous tree level  case.  To this end we will  suppose that the loop-contributions  are  dominant in the conversion 
process as well as in the dispersion relations.  In such  a situation,  the hidden mass term  can be ignored,  the original  $\rm U(1)\times U(1)$-symmetry  is   preserved and the 
Lagrangian density in  Eq.~(\ref{startingLagrangianII}) becomes the starting point of further considerations. 

We shall assume throughout that the charged particles involved in Eq.~(\ref{startingLagrangianII}) are  coupled  to a  circularly polarized monochromatic plane-wave 
\begin{eqnarray}
\mathscr{A}^{\mu}(x)=\mathpzc{a}_1^{\mu}\cos(\varkappa x)+\mathpzc{a}_2^{\mu}\sin(\varkappa x)
\label{externalF}
\end{eqnarray} by  the standard minimal-coupling scheme so that the following  Lagrangian density 
\begin{equation}
\mathpzc{L}_{\mathrm{ext}}=\chi e_h j_h^\mu \mathscr{A}_\mu \label{secondstartingLagrangianII}
\end{equation} must be added to $\mathpzc{L}$ [Eq.~(\ref{startingLagrangianII})]. The external field in Eq.~(\ref{externalF})  is chosen in the  Lorenz  gauge  $\partial\mathscr{A}=0$. 
Therefore, the wave four-vector  $\varkappa^{\mu}=(\varkappa^0,\pmb{\varkappa})$   and  the  constant vectors  along the  polarization directions  $\mathpzc{a}_i^\mu$   (with  $i=1,2$) 
satisfy the  relations  $\varkappa \mathpzc{a}_i=0$, $\varkappa^2=0$, and  $\mathpzc{a}_1^2=\mathpzc{a}_2^2\equiv\mathpzc{a}^2$. 

\begin{figure}
\includegraphics[width=6in]{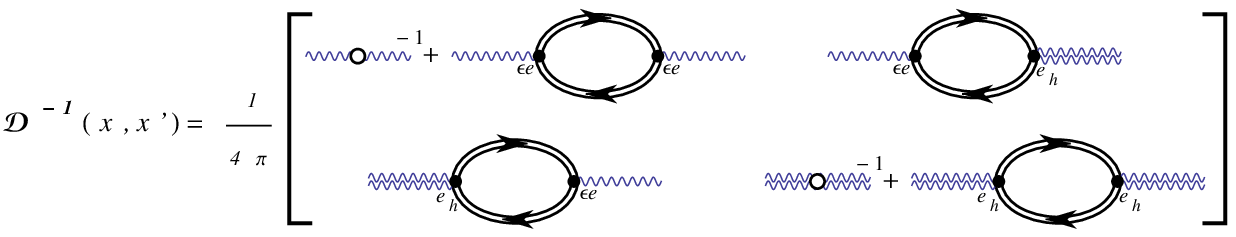}
\caption{\label{fig:mb001}Pictorial representation of the inverse Green's function $\pmb{\mathpzc{D}}^{-1}(x,x^\prime)$ in the one-loop approximation.  The double lines represent the propagator of  MCPs 
including the full interaction with the external field. A  single wavy line denotes the amputated leg  corresponding to a  small-amplitude electromagnetic wave. Conversely,  a double wavy 
line refers to the  amputated leg associated with a  hidden-photon field.  The unperturbated propagators--inverses of  the leading order operators  in the 
right-hand-side of Eq.~(\ref{inverseppropgator})--are indicated  by open blobs.}
\end{figure}

The  equations of motion that follow  from the combination of  Eqs.~(\ref{startingLagrangianII}) 
and (\ref{secondstartingLagrangianII}) can be  used to determine  the effective action  as it follows  from a  Legendre transform of the generating 
functional of the connected Green's  functions.  The   integro-differential ansatz  which allows us to reach this aim  is  known in the literature \cite{Rivers,fradkin,Alkofer:2000wg,VillalbaChavez:2008dv}.
Its application to the problem  under 
consideration leads  to express the  gauge sector of the generating functional of one-particle irreducible Feynman graphs in the following form
\begin{equation}\label{effectiveaction}
\Gamma[\pmb{\Phi}]=\frac{1}{2}\int d^4x\ d^4x^\prime\ \pmb{\Phi}^\mathrm{T}(x)\pmb{\mathpzc{D}}^{-1}(x,x^\prime)\pmb{\Phi}(x^\prime)+\ldots,
\end{equation}where the abbreviation   $+\ldots$ stands for higher order terms in  the small-amplitude  gauge fields. The  inverse Green's function $\pmb{\mathpzc{D}}^{-1}(x,x^\prime)$ 
 and the flavor field $\pmb{\Phi}(x)$ involved in Eq.~(\ref{effectiveaction})  are given by 
\begin{equation}\label{eqIII}
\pmb{\mathpzc{D}}^{-1}(x,x^\prime)\equiv\left[
\begin{array}{cc}
\mathscr{D}_{\mu\nu;a}^{-1}(x,x^\prime)&  \frac{1}{4\pi}\Pi_{\mu\nu;o}(x,x^\prime)\\ \\
 \frac{1}{4\pi}\Pi_{\mu\nu;o}(x,x^\prime)& \mathscr{D}_{\mu\nu;w}^{-1}(x,x^\prime)
\end{array}\right]\qquad\mathrm{and}\qquad \pmb{\Phi}(x)=\left[\begin{array}{c}a_\mu(x)\\ \\ w_\mu(x)\end{array}\right],
\end{equation} respectively. The diagonal components in  $\pmb{\mathpzc{D}}^{-1}(x,x^\prime)$   are the respective   two-points  proper correlation functions associated with the 
fields  $a_{\mu}(x)$ and $w_{\mu}(x)$. Explicitly, 
\begin{eqnarray}\label{inverseppropgator}
\mathscr{D}_{\mu\nu;i}^{-1}(x,x^\prime)=\frac{1}{4\pi}\left(\square \mathpzc{g}_{\mu\nu}-\partial_\mu\partial_\nu\right)\delta^4(x-x^\prime)+\frac{1}{4\pi}\Pi_{\mu\nu;i}(x,x^\prime),\quad i=a,w 
\end{eqnarray}where $\square\equiv\partial_\mu\partial^\mu=\partial^2/\partial t^2-\nabla^2$ and  the metric tensor reads $\mathpzc{g}_{\mu\nu}=\mathrm{diag}(+1,-1,-1,-1)$.  
The off-diagonal terms in Eq.~(\ref{eqIII})  $ \frac{1}{4\pi}\Pi_{\mu\nu;o}(x,x^\prime)$ and  the last contribution embedded in  the expression above, i. e., 
$ \frac{1}{4\pi}\Pi_{\mu\nu;i}(x,x^\prime)$ with $i=a,w$ enconde the analytic structures  of the  two-point  irreducible Feynman  diagrams. While the latter  
terms  are responsible for  pure  scattering  processes,\footnote{For instance,  $\Pi_{\mu\nu;a}$ contributes  to    photon-photon scattering.}  the former drives 
the  photon-paraphoton  oscillations. In such a case, the physical propagating modes are certain  mixtures of photon and hidden-photon states resulting 
from the diagonalization of $\pmb{\mathpzc{D}}^{-1}(x,x^\prime)$.   The described forms  of  interactions  are  determined  by the causal Feynman propagator  of 
the  minicharged carriers in the external field [Furry  picture]. Because of this fact,  a dependence  on the strong  electromagnetic background  is  introduced.  

Obviously, in the one-loop  approximation, the dressed vertices are  reduced to  those involved  in the initial Lagrangian density [Eq.~(\ref{startingLagrangianII})]. 
As a consequence, the inverse Green's  function $\pmb{\mathpzc{D}}^{-1}(x,x^\prime)$ acquires a simple diagramatic representation [see Fig.~\ref{fig:mb001}] and  
the  tensors  driving  the photon-paraphoton oscillations and the  respective  paraphoton scattering process   become proportional to 
$\Pi_{\mu\nu;a}(x,x^\prime)\equiv \Pi_{\mu\nu}(x,x^\prime)$. Explicitly, 
\begin{equation}\label{onelooprelation}
\Pi_{\mu\nu;o}(x,x^\prime)=-\frac{1}{\chi}\Pi_{\mu\nu}(x,x^\prime)\quad\mathrm{and}\quad \Pi_{\mu\nu;w}(x,x^\prime)=\frac{1}{\chi^2}\Pi_{\mu\nu}(x,x^\prime),
\end{equation}where Eq.~(\ref{chargesrelation}) has been used. The  analytical properties  of $\Pi_{\mu\nu}(x,x^\prime)$  do not differ from  the  vacuum polarization tensor  
that arises  in a pure QED context. Hence, an appropriate replacement  of  the  electron  parameters $(e,\ m)$  by the respective quantities associated with an MCP $(\mathpzc{q}_\epsilon,\ m_\epsilon)$ 
is enough for acquiring the necessary  insights on  the structural nature of  $\Pi_{\mu\nu}(x,x^\prime)$, and in the related   forms of interactions  [Eq.~(\ref{onelooprelation})].  

We conclude this subsection by obtaining the  Dyson-Schwinger equations \cite{Rivers,fradkin,Alkofer:2000wg}  from 
the quadratic part  of our effective action  [Eq.~(\ref{effectiveaction})]. In momentum space they read
\begin{eqnarray}
&&k^2 a_\mu(k)-\int \frac{d^4k^\prime}{(2\pi)^4} \Pi_{\mu\nu}(k,k^\prime) a^\nu(k^\prime)+\frac{1}{\chi}\int \frac{d^4k^\prime}{(2\pi)^4} \Pi_{\mu\nu}(k,k^\prime) w^\nu(k^\prime)=0,\label{main1}\\
&& k^2 w_\mu(k)-\frac{1}{\chi^2}\int\frac{d^4k^\prime}{(2\pi)^4} \Pi_{\mu\nu}(k,k^\prime)w^\nu(k^\prime)+\frac{1}{\chi}\int \frac{d^4k^\prime}{(2\pi)^4} \Pi_{\mu\nu}(k,k^\prime)a^\nu(k^\prime)=0,\label{main2}
\end{eqnarray} provided that both Abelian fields  are chosen in the Lorenz gauge $k^\mu a_\mu=0$,  $k^\mu w_\mu=0$. Note that so far, we have  not made use of the precise form of the external 
wave [Eq.~(\ref{externalF})] and therefore, the formulae above  apply whatever be the nature of the background electromagnetic field. Furthermore, by ignoring the terms proportional 
to $\sim1/\chi$, one can analyze a  model in which MCPs exist  without the  occurence of both:  the   kinetic-mixing term  and the hidden-photon field. Our study  intends  
to establish  comparisons between this pure MCPs scenario and the full model  described by Eq.~(\ref{main1}) and (\ref{main2}).

\subsection{Absorption and dispersion of small-amplitude waves \label{ADsaw}}

In the field of a circularly polarized wave [Eq.~(\ref{externalF})],   the  polarization tensor in momentum space splits  into two relevant  terms:
\begin{eqnarray}\label{ptcircular}
\Pi^{\mu\nu}(k,k^\prime)=(2\pi)^4\delta^4(k-k^\prime)\Pi^{\mu\nu}_0(k^\prime)+\sum_{j=+,-}(2\pi)^4\delta^4(k-k^\prime+2j\varkappa)\Pi^{\mu\nu}_j(k^\prime)
\end{eqnarray}out of which  the inelastic   contribution--second term in Eq.~(\ref{ptcircular})--describes scattering process  characterized by  
the simultaneous  emission or  absorption of photons of the high-intensity laser wave upon the scattering event. The precise structure of $\Pi^{\mu\nu}_\pm(k^\prime)$ 
is not relevant for what  follows. However, the part responsible for the elastic process--first term in Eq.~(\ref{ptcircular})--deserves to be explained  in 
some detail. This contribution is diagonalizable by using a vector basis that   manifests   the underlying  invariance properties  of the vacuum
\begin{eqnarray}\label{elasticpart}
\Pi^{\mu\nu}_0(k)=\sum_{i=+,-} \pi_i\Lambda^\mu_i\Lambda^{\nu*}_i
\end{eqnarray}because the two  relevant contributions are determined by transverse eigenstates  of opposite  helicities  $k_\mu\Lambda_\pm^\mu =0$,  subject to  
the normalization conditions $\Lambda_+\Lambda_-=-1$ with  $\Lambda_\pm\Lambda_\pm=0$ and $\Lambda^*_\pm=\Lambda_\mp$. Formally, there should occur  two additional 
terms  in  the diagonal expansion in Eq.~(\ref{elasticpart}). One of these ommited contributions  turns out to be  longitudinal by construction 
$\sim k^\mu k^\nu$,  but  owing to  the  gauge invariance property of the polarization  tensor $\Pi_0^{\mu\nu}k_\nu=0$,   its corresponding  eigenvalue vanishes 
identically. It is worth mentioning  that  the previous statement is independent of any approximation used  in the calculation of $\Pi_{\mu\nu}$. The 
remaining disregarded term is originally proportional to a transverse four-vector  $\Lambda^\mu_3\sim \varkappa^\mu k^2-k^{\mu} (k\varkappa)$,  its eigenvalue 
being proportional to  $k^2$ in the one-loop approximation. The latter  leads to a trivial dispersion equation $k^2=0$ in which case  $\Lambda^\mu_3\sim k_\mu$   
becomes a longitudinal vector that   cannot be associated with a physical propagating mode \cite{baier,VillalbaChavez:2012bb,Villalba-Chavez:2013gma}.

The relevant  eigenvalues $\pi_\pm=-(\pi_3\pm i\pi_1)$ are unwieldy complex functions determined by the form factors $\pi_{1,3}$ introduced  by Ba\u{\i}er, Mil'shte\u{\i}n 
and Strakhovenko in Ref.~\cite{baier}. Accordingly, they  are  represented  as  twofold parametric integrals which, in general,  cannot be evaluated  analytically.  
In fact, when  the polarization effects  are  tiny  corrections to  the free photon dispersion equation [$k^2=\mathpzc{w}^2-\pmb{k}^2\simeq0$], the inelastic contributions 
are  strongly suppressed  by energy-momentum  conservation  \footnote{The partial  production rates associated with the generation of an inelastic scattered wave accompanied 
by a flip of polarization can   be read off from the modulus square of the  amplitudes that follow from the last term in  Eq.~(\ref{ptcircular}) after integration over the final 
momentum space.  Because of the associated  Dirac deltas $\delta^4(k-k^\prime\pm2\varkappa)$,  the last  operation leads to partial rates proportional to  
$\propto \delta(\vert\pmb{k}\vert-\vert\pmb{k}\pm2\pmb{\varkappa}\vert\pm2\varkappa_0)$.  The appearance of these final  Dirac deltas  is intrinsically connected with the 
monochromaticity   of the strong  wave. The  energetic  balances imposed by them  cannot be fulfilled in general,  leading to  vanishing rates. For finite pulses the 
situation is somewhat different (details can be found in Refs.~\cite{affleck,Villalba-Chavez:2013gma,Dinu:2013gaa}). } and one finds that
\begin{equation}\label{formfactors}
\pi_\pm(n_*,\xi_\epsilon)=\frac{\alpha_\epsilon}{2\pi} m_\epsilon^2\int_{-1}^{1}dv \int_0^\infty \frac{d\rho}{\rho}\Omega_\pm \exp\left\{-\frac{2i\rho n_*}{(1+\xi_\epsilon^2)(1-v^2)}\left[1+2A\xi_\epsilon^2\right]\right\}.
\end{equation} While $\alpha_\epsilon\equiv \epsilon^2 e^2=\epsilon^2/137$ denotes the fine structure constant relative to the MCPs,  
\begin{equation}
 n_*=2\frac{m_\epsilon^2(1+\xi_\epsilon^2) }{k\varkappa}\qquad \mathrm{with}\qquad\xi_\epsilon^2=-\frac{\epsilon^2 e^2\mathpzc{a}^2}{m_\epsilon^2}\label{Cpar}
\end{equation}refers to  the threshold parameter for the photo-production of a  $\mathpzc{q}_\epsilon^+\mathpzc{q}_\epsilon^--$pair.   

The difference between each eigenvalue [Eq.~(\ref{formfactors})] is originally introduced  by the functions $\Omega_\pm$:
\begin{eqnarray}
&&\Omega_\pm=2\xi_\epsilon^2\frac{1+v^2}{1-v^2}\left[\sin^2(\rho)\pm2i\rho A_0\right]-1+\exp(iy).\label{g3fermion}
\end{eqnarray} Other functions and parameters  contained  in  the above  expressions,   are given by
\begin{eqnarray}
\begin{array}{c}\displaystyle
A=\frac{1}{2}\left[1-\frac{\sin^2(\rho)}{\rho^2}\right],\quad  A_0=\frac{1}{2}\left[\frac{\sin^2(\rho)}{\rho^2}-\frac{\sin(2\rho)}{2\rho}\right],\quad y=\frac{4 n_* \xi_\epsilon^2 \rho A}{(1+\xi_\epsilon^2)(1-v^2)}.
\end{array}
\label{parameters}
\end{eqnarray} With the change of variable $(1-v^2)^{-1}\to \frac{1}{2}[\cosh(t)+1]$ and the succeding identification of the  Hankel function of second kind 
$\mathrm{H}_{\nu}^{(2)}(z)=\frac{2i}{\pi}\exp[\frac{i}{2}\pi\nu]\int_0^{\infty} dt \exp[-iz\cosh(t)]\cosh(\nu t)$, the 
variable $v$ is integrated out and the $\Pi_0^{\mu\nu}-$eigenvalues acquire the compact  structures 
\begin{equation}\label{eigenvalues}
\pi_\pm(n_*,\xi_\epsilon)=\frac{1}{2} \alpha_\epsilon m_\epsilon^2 \int_0^\infty \frac{d\rho}{\rho}\Upsilon_\pm \exp\left(-i\eta\right),
\end{equation}where $\eta\equiv\rho n_*\left(1+2\xi_\epsilon^2A\right)/(1+\xi_\epsilon^2)$  and  the functions $\Upsilon_\pm$ read
\begin{eqnarray}
&&\Upsilon_\pm=\left\{\eta\left[\mathrm{H}_0^{(2)}(\eta)+i\mathrm{H}_1^{(2)}(\eta)\right]-i\mathrm{H}_0^{(2)}(\eta)\right\}\left[2\xi_\epsilon^2\sin^2(\rho)\pm 4\xi_\epsilon^2 \rho A_0\right]+\eta\left[\mathrm{H}_0^{(2)}(\eta)+i\mathrm{H}_1^{(2)}(\eta)\right]\nonumber\\
&&\qquad\qquad-\rho n_* \left[\mathrm{H}_0^{(2)}(\rho n_*)+i\mathrm{H}_1^{(2)}(\rho n_*)\right]\exp\left[\frac{2i\xi_\epsilon^2 \rho n_* A }{1+\xi_\epsilon^2}\right].\label{integrand2}
\end{eqnarray} 

In order to pursue our analysis we  seek the  flavor-like solutions of the problem  as  superpositions of tranverse eigenwaves with opposite helicities 
\begin{equation}\label{chiralityexpansio}
a^{\mu}(k)=f_+(k)\Lambda^\mu_++f_-(k)\Lambda^\mu_-\quad \mathrm{and}\quad  w^{\mu}(k)=g_+(k)\Lambda^\mu_++g_-(k)\Lambda^\mu_-. 
\end{equation}Eqs.~(\ref{ptcircular})-(\ref{chiralityexpansio}) are then  inserted into Eq.~(\ref{main1}) and (\ref{main2}).  
As a consequence, the problem defined by the latter formulae   splits  into  two  eigenproblems; each one associated with a unique value of helicity  as one  
can expect from the   angular-momentum conservation 
\begin{eqnarray}
\label{dispr1}
\left[
\begin{array}{cc}
k^2-\pi_\pm & \frac{1}{\chi}\pi_\pm\\
\frac{1}{\chi}\pi_\pm& k^2-\frac{1}{\chi^2}\pi_\pm
\end{array}\right]\left[\begin{array}{c} f_\pm(k)\\ g_\pm(k)
\end{array}\right]=0.
\end{eqnarray} Next, the leading terms in the diagonal elements   are  linearized according to the rule  $k^2\simeq 2\omega_{\pmb{k}}(\mathpzc{w}-\omega_{\pmb{k}})$ with $\omega_{\pmb{k}}\equiv\vert\pmb{k}\vert$, 
which  is  equivalent to  reducing  the order in the differential versions of the  equations of motion [Eq.~(\ref{main1}) and (\ref{main2})] (see Sec.~\ref{CPPO}). The dispersion  relations  are then 
established   by setting the  determinants of the  resulting  matrices  to zero. Explicitly, we obtain
\begin{equation}\label{disperisonlaws}
\mathpzc{w}_\pm^{(\gamma)}=\omega_{\pmb{k}}+\frac{\pi_\pm}{2\omega_{\pmb{k}}} \quad \mathrm{and}\quad \mathpzc{w}_\pm^{(\gamma^\prime)}=\omega_{\pmb{k}}+\frac{\pi_\pm}{2\chi^2\omega_{\pmb{k}}}.
\end{equation}Hereafter the symbols  $\gamma$ and  $\gamma^\prime$ label the dispersion laws  associated with the physical modes of visible and  hidden-photon fields, respectively.  
Note that  the contributions resulting from the off-diagonal  terms  have been ignored  because they  provide  corrections smaller by a factor  $\sim(\epsilon e)^2e_h^2$.  
 
Owing to the non-hermiticity of the vacuum  polarization tensor, its  eigenvalues   can be decomposed in terms of their  real and imaginary parts  
$\pi_\pm=\mathrm{Re}\ \pi_\pm+i\ \mathrm{Im}\ \pi_\pm$. This fact renders  the dispersion relations [Eq.~(\ref{disperisonlaws})]  complex functions too with  
$\mathpzc{w}_\pm=\mathrm{Re}\ \mathpzc{w}_\pm+i\ \mathrm{Im}\ \mathpzc{w}_\pm$. As a consequece, we can  define  the vacuum refractive indices  
$n_\pm=\vert\pmb{k}\vert/\mathrm{Re}\ \mathpzc{w}_\pm$ and the corresponding  absorption  coefficients $\kappa_\pm\equiv-\mathrm{Im}\ \mathpzc{w}_\pm$ 
associated with each propagating mode of the respective   Abelian fields:
\begin{eqnarray}
\begin{array}{cc}\displaystyle
n_\pm-1=-\left.\frac{\mathrm{Re}\ \pi_\pm}{2 \omega_{\pmb{k}}^2}\right\vert_{k^2=0}=\left.\frac{\mathrm{Re}\ \pi_3\mp\mathrm{Im}\ \pi_1}{2\omega_{\pmb{k}}^2}\right\vert_{k^2=0},\quad&\displaystyle n_\pm^{(\gamma^\prime)}-1=\frac{1}{\chi^2}\left(n_\pm-1\right)\\ \\
\displaystyle
\kappa_\pm=-\left.\frac{\mathrm{Im}\ \pi_\pm}{2\omega_{\pmb{k}}}\right\vert_{k^2=0}=\left.\frac{\mathrm{Im}\ \pi_3\pm\mathrm{Re}\ \pi_1}{2 \omega_{\pmb{k}}}\right\vert_{k^2=0},\qquad &\displaystyle \kappa_\pm^{(\gamma^\prime)}=\frac{1}{\chi^2}\kappa_\pm,
\end{array}\label{refractioninde}
\end{eqnarray} where  the decomposition $\pi_\pm=-(\pi_3\pm i\pi_1)$  has been used. Observe that the terms proportional to the real and imaginary parts  of  
$\pi_1$  determine the   degree of dichroism and  birefringence of the vacuum   polarized by the external laser  wave [Eq.~(\ref{externalF})], respectively.  
The former phenomenon is closely associated with the different amount of pairs of  MCPs produced by each propagating mode.  In the field of the wave, the production 
thresholds are determined by the condition $n\geqslant n_*$ [see Eq.~(\ref{Cpar})] with $n$  denoting the  minimal  number  of photons from the strong wave that kinematically allows 
the multiphoton process $k+n\varkappa\to \mathpzc{q}_\epsilon^++ \mathpzc{q}_\epsilon^-$.  Note that the previous relation leads to a condition $m_\epsilon\leqslant m_n$ 
which depends on  the threshold mass
\begin{equation}
m_n\equiv\sqrt{\frac{1}{2} nk\varkappa-\epsilon^2m^2\xi^2},\label{thresholdmass}
\end{equation}with  $\xi^2=-e^2\mathpzc{a}^2/m^2$ refering  to  the  usual laser intensity  parameter with $m$ being the electron mass. Clearly, Eq.~(\ref{thresholdmass}) provides real 
threshold masses  whenever the condition $n\kappa k/(2m^2\xi^2)\geqslant\epsilon^2$ is satisfied.  Close to the lowest thresholds of  pair production of MCPs  [$n_*\sim1$]  the  
chiral  birefringence and dichroism  properties of the vacuum are predicted to be considerably more  pronounced  than in the cases asymptotically far from it [$n_*\to \infty$ and $n_*\to0$], at 
which the vacuum behaves like a nonabsorbing isotropic medium  \cite{Villalba-Chavez:2013gma}.  In the following our attention is focused on the simple cases in which one or 
two  photons from the strong wave [$n=1,2$] are absorbed, i. e., the limits of the two $(k+\varkappa)$ and three-photon $(k+2\varkappa)$ reactions. Contributions of higher thresholds  $[n>2]$ are beyond 
the scope of this work. This is partially motivated by the fact that for $\xi_\epsilon<1$  the photo-production rate at higher thresholds  scales as  
$\mathpzc{R}_{\ n}\propto \xi_\epsilon^{2n}$ \cite{Di_Piazza_2012,Greiner},  which provides an evidence that the production of $\mathpzc{q}_\epsilon^+\mathpzc{q}_\epsilon^-$ 
pairs by means of  the absorption of  several photons from the external  wave is less likely to occur.

\subsection{Spectral decomposition  at  $\xi_\epsilon<1$ \label{spectraldecompos}}

The integrands which define the  $\Pi_0^{\mu\nu}-$eigenvalues are functions of the variable  $\eta=n_*\rho(1-\Delta)$ with $\Delta=\frac{\xi_\epsilon^2}{(1+\xi_\epsilon^2)}\frac{\sin^2(\rho)}{\rho^2}$.
In the region of interest [$\xi_\epsilon<1$], this  factor  is much   smaller  than  unity and   the  respective Taylor  expansions  of the integrands  lead to a sum of 
contributions  coming from the threshold points.  To show this we first  consider the lowest order terms  with respect to $\Delta$. Once the integral representations  
of the remainig  Hankel functions $\mathrm{H}_{0,1}^{(2)}(\rho n_*)$ are  used [see below Eq.~(\ref{parameters})], we  can exchange the order of integration and first integrate 
over $\rho$.\footnote{The procedure outlined here shares several similarities with the method applied by the authors in \cite{Villalba-Chavez:2013gma}, particularly in 
those issues associated with the integrations over the  variable $\rho$. The  reader interested in the details of such operations may find it  helpful  to refer  to the aforementioned 
reference.}   The real parts of the resulting integrands  turn out to be  discontinous functions at $n_*=1$ and  determine the leading order 
terms of the  absorption coefficients.  After having  integrated out the remaining integration variable, they read 
\begin{eqnarray}
&&\kappa_{+,1}=\frac{\alpha_\epsilon m^2_\epsilon \xi_\epsilon^2}{4\omega_{\pmb{k}}}\left\{\frac{1-\mathpzc{v}_1^4}{2(1+\xi_\epsilon^2)}\ln\left(\frac{1+\mathpzc{v}_1}{1-\mathpzc{v}_1}\right)+2\mathpzc{v}_1\left(1-\frac{1-\mathpzc{v}_1^2}{2(1+\xi_\epsilon^2)}\right)\right\}\Theta[\mathpzc{v}_1^2],\label{fermionpositivekappa} \\ 
&&\kappa_{-,1}=\frac{\alpha_\epsilon m^2_\epsilon \xi_\epsilon^2}{4\omega_{\pmb{k}}}\left\{\left(2+\frac{1-\mathpzc{v}_1^4}{2(1+\xi_\epsilon^2)}\right)\ln\left(\frac{1+\mathpzc{v}_1}{1-\mathpzc{v}_1}\right)-4\mathpzc{v}_1\left(1+\frac{1-\mathpzc{v}_1^2}{4(1+\xi_\epsilon^2)}\right)\right\}\Theta[\mathpzc{v}_1^2].\label{fermionnegativekappa}
\end{eqnarray} Here  $\Theta[x]$ represents  the unit step  function, whereas   $\mathpzc{v}_1=(1-n_*)^{\nicefrac{1}{2}}$ is closely connected to  the relative speed of  the  final particle states 
when only one   photon of the intense  laser wave  has been  absorbed;\footnote{In the center--of--mass frame, when  $n$ photons from the laser field are absorbed, the relative speed between  the final particles   is 
given by $\vert\pmb{\mathpzc{v}}_{\mathrm{rel}}\vert=\vert\pmb{v}_{\mathpzc{q}_\epsilon^-}-\pmb{v}_{\mathpzc{q}_\epsilon^+}\vert=2\mathpzc{v}_n$ with $\mathpzc{v}_n=(1-n_*/n)^{\nicefrac{1}{2}}$.}  hence the use of  the lower index $1$.  We emphasize that  Eqs.~(\ref{fermionpositivekappa})-(\ref{fermionnegativekappa})  
provide  nonvanishing contributions whenever the MCP mass $m_\epsilon$ is smaller or equal to the first  threshold mass 
$m_1=\left(k\varkappa/2-\epsilon^2m^2\xi^2\right)^{\nicefrac{1}{2}}$, corresponding to $n_*\leqslant1$.  

In contrast to the previous case, the imaginary parts of the $\pi_\pm$-integrands  are continuous functions which define the leading terms  of  the vacuum refractive indices. The  explicit expressions of these optical entities are difficult to obtain.  Asymptotic
expressions for $n_*\ll1$ and $n_*\approx1$ can be found but both cases have restricted  validity   in comparison to Eqs.~(\ref{fermionpositivekappa})-(\ref{fermionnegativekappa}). 
Because  of this reason, we have opted to express them  as  parametric integrals:
\begin{eqnarray}
&& n_\pm-1\simeq\pm\frac{\alpha_\epsilon m_\epsilon^2  \xi_\epsilon^2}{2\pi\omega_{\pmb{k}}^2} \int_{0}^1dv\left\{\left[1-\frac{2\varrho}{n_*}\left(1\mp \frac{n_*}{1+\xi_\epsilon^2}\right)\right]\ln\left(\frac{1+\varrho}{\vert1-\varrho\vert}\right)^{\nicefrac{1}{2}}\right.\nonumber\\ 
&&\qquad\qquad\left.\mp\left[1-\frac{2\varrho}{n_*}(1\mp n_*)+\frac{2\varrho^2}{1+\xi_{\epsilon}^2}\left(1\mp\frac{2(1+\xi_\epsilon^2)}{n_*}\right)\right]\ln\left(\frac{\vert\varrho\vert}{\sqrt{\vert1-\varrho^2\vert}}\right)\right\}\label{interpi3},
\end{eqnarray} where $\varrho\equiv\varrho(v,n_*)=n_*(1-v^2)^{-1}$ is a function of both the integration variable $v$ and the threshold parameter $n_*$.

Equation (\ref{interpi3}) deserves further analysis.  To simplify it,  we only  keep the quadratic proportionality  on $\xi_\epsilon$  in $n_\pm-1$  focusing 
on the Born approximation [$\xi_\epsilon^2\ll1$]. As a consequence, the behavior of the integrand of $n_\pm$ as  $v\to 0$ 
[$\varrho\to n_*$] turns out to be
\begin{eqnarray}
&&\xrightarrow{v\to0}-\left[1\mp2 n_*\right]\ln\left(\frac{1+n_*}{1-n_*}\right)^{\nicefrac{1}{2}}\pm\left[1\pm 2n_*-2n_*^2\right]\ln\left(\frac{n_*}{\sqrt{1-n_*^2}}\right).
\end{eqnarray}Conversely, when $v\to1$ the function $\varrho\to\infty$ and the integrand  associated with $n_\pm$ tends to $\pm1$. The situation  is different 
at  $v\to \sqrt{1-n_*}<1$ with $n_*\in(0,1)$, i. e.,  $\varrho\to 1$, since the derivation of  $\pi_\pm$  relies on integrals tabulated in 
Ref.~\cite{Gradshteyn},  which  apply whenever $\varrho\neq1$, for further details see \cite{Villalba-Chavez:2013gma}. Therefore, this singularity is  actually not   
reached  and, in  a neighborhood of this point  the integrand  of $n_\pm$ behaves  as  $\sim \frac{1}{2}(1\pm1\mp\frac{2}{n_*})\ln\vert1-\varrho\vert$. 
Note that, due to the exclusion of the aforementioned point, the integral over $v$ in Eq.~(\ref{interpi3})  must be understood  as a Cauchy principal value. However, 
if $n_*\gtrsim1$ the  square root of $1-n_*$  is an imaginary quantity. As a consequence,  there is  no singularity  within the integration region $0\leqslant v\leqslant1$ and--in contrast to 
Eqs.~(\ref{fermionpositivekappa})-(\ref{fermionnegativekappa})--the leading order  terms  of the vacuum refractive indices turn out to be dominant,  even when the 
contributions resulting from  higher thresholds are taken into account.

Of particular interest for us are  the corrections to  Eqs.~(\ref{fermionpositivekappa})-(\ref{fermionnegativekappa}) which result  from  the absorption of  two photons from the strong 
wave. To determine them  we  assume that  $1< n_*\leq2$, so that the $\xi^4_\epsilon-$correction due to  two-photon reaction $(k+\varkappa)$ is  excluded.  With this detail  in 
mind we go one step further in the $\Delta$-expansion of Eqs.~(\ref{eigenvalues})-(\ref{integrand2}). We then  find that the real parts of the integrands associated with the three-photon 
reaction $(k+2\varkappa)$ are, in general, discontinuous functions at the  point where $n_*=2$ . Following the procedure  described  above we obtain 
\begin{eqnarray}\label{absorptioncoefficient2order}
\kappa_{\pm,2}=\frac{\alpha_\epsilon m^2_\epsilon \xi_\epsilon^4}{4\omega_{\pmb{k}}(1+\xi_\epsilon^2)}\left[\mathpzc{F}_1(\mathpzc{v}_2)+2\frac{1-\mathpzc{v}_2^2}{1+\xi_\epsilon^2}\mathpzc{F}_2(\mathpzc{v}_2)\pm \mathpzc{F}_3(\mathpzc{v}_2)\right]\Theta[\mathpzc{v}_2^2]
\end{eqnarray}where $\mathpzc{v}_2=(1-n_*/2)^{\nicefrac{1}{2}}$ defines the relative speed between the produced minicharges [see footnote~$4$], whereas the functions $\mathpzc{F}_i(\mathpzc{v}_2)$ 
with $i=1,2,3$ are given by 
\begin{eqnarray}\label{F1fermion}
&&\mathpzc{F}_1(\mathpzc{v}_2)=\mathpzc{v}_2\left(1+\mathpzc{v}_2^2\right)-\left(1-\mathpzc{v}_2^2\right)^2\arctanh(\mathpzc{v}_2),\\ 
&&\mathpzc{F}_2(\mathpzc{v}_2)=\frac{1}{12}\mathpzc{v}_2\left(15\mathpzc{v}_2^4-4\mathpzc{v}_2^2-3\right)+\frac{1}{4}\left(1+\mathpzc{v}_2^2+3\mathpzc{v}_2^4-5\mathpzc{v}_2^6\right)\arctanh(\mathpzc{v}_2),\\
&&\mathpzc{F}_3(\mathpzc{v}_2)=-\frac{1}{3}\mathpzc{v}_2\left(6\mathpzc{v}_2^4-7\mathpzc{v}_2^2+3\right)+\left(1-3\mathpzc{v}_2^4+2\mathpzc{v}_2^6\right)\arctanh(\mathpzc{v}_2),
\end{eqnarray}
The above formulae allow us to determine  the asymptotic  expression of $\kappa_{\pm,2}$ as $n_*\to 2^-$, i. e., when the particles are created in the 
center-of-mass frame almost at rest [$\mathpzc{v}_2\sim0$].  In this limit the functions $\mathpzc{F}_i(\mathpzc{v}_2)$ are dominated by  cubic dependences 
on $\mathpzc{v}_2$ and the absorption coefficients  approach to  $\kappa_{\pm,2}\approx\alpha_\epsilon m^2_\epsilon \xi_\epsilon^4 \mathpzc{v}_2^3(8\mp1)/[12\omega_{\pmb{k}}(1+\xi_\epsilon^2)]$. 
Conversely, when $n_*\to1^+$, i. e., [$\mathpzc{v}_2\to\nicefrac{1}{\sqrt{2}}$],  we find  the asymptotes  
$\kappa_{\pm,2}\approx\alpha_\epsilon m^2_\epsilon \xi_\epsilon^4 (0.4\mp0.1)/[4\omega_{\pmb{k}}]$, provided 
the condition $\xi_\epsilon\ll1$ holds. The asymptotic behavior of $\kappa_{\pm,1}$ was derived  previously and can be found in Ref.~\cite{Villalba-Chavez:2013gma}

Finally,  we recall that  the imaginary part of the polarization tensor is associated with the production  rate  of a $\mathpzc{q}_\epsilon^+\mathpzc{q}_\epsilon^-$pair  
through the optical theorem. Within the  accuracy to the second order with respect to the radiative corrections, the total creation rate $\mathpzc{R}$  of a 
$\mathpzc{q}_\epsilon^+ \mathpzc{q}_\epsilon^-$ pair from a photon--averaged  over  the polarization states $\Lambda_\pm^\mu$--is fully determined by the absorption 
coefficients~\cite{Villalba-Chavez:2013gma}.  Explicit polarization operator approaches to the rate  associated  with the two-photon reaction  may be found in  separate papers (see Refs. \cite{VillalbaChavez:2012bb} and \cite{baierbook}). 
In  the  limit of $\xi_\epsilon \ll1$  it was obtained that $\mathpzc{R}_{\ 1}\propto\xi^2_\epsilon$. When inspecting  Eq.~(\ref{absorptioncoefficient2order}) one can easily 
establish that  the  average  rate  for producing a $\mathpzc{q}_\epsilon^+\mathpzc{q}_\epsilon^-$pair in  a three-photon reaction is $\mathpzc{R}_{\ 2}\propto\xi_\epsilon^4$, a fact which 
verifies the last comment in Sec.~\ref{ADsaw}.

\subsection{Conversion probability and  polarimetric  observables \label{CPPO}}

The linearization  used in the derivation of Eq.~(\ref{disperisonlaws}) is also a convenient  simplification  for solving  the  initial system  
of differential equations [Eqs.~(\ref{main1}) and (\ref{main2})]. It turns out to be  appropriate  to seek for solutions  in the  form of plane 
waves $\sim e^{i\pmb{k}\cdot\pmb{x}-i\omega_{\pmb{k}}t}$. This fact   allows us to approximate  the   Laplacian 
involved in the equations of motion [see below Eq.~(\ref{inverseppropgator})]  by a first order differential operator according to the rule  
$\partial^2/\partial t^2+\pmb{k}^2=\left(i\partial/\partial t+\vert\pmb{k}\vert\right)\left(-i\partial/\partial t+\vert\pmb{k}\vert\right) \simeq2\omega_{\pmb{k}}\left(-i\partial/\partial t+\omega_{\pmb{k}}\right)$.
As a consequece,  the  boundary conditions on the derivatives  of  both gauge fields can be ignored and the problem under consideration  
reduces to solve  the  equation 
\begin{equation}\label{reduceddispersionequation}
 -i\frac{\partial}{\partial t}\left[\begin{array}{c} f_\pm(\pmb{k},t)\\ \\ g_\pm(\pmb{k},t)\end{array}\right]=
\left[
\begin{array}{cc}
\mathpzc{w}_\pm^{(\gamma)} & \frac{1}{2\chi\omega_{\pmb{k}}}\pi_\pm\\  \\
\frac{1}{2\chi\omega_{\pmb{k}}}\pi_\pm& \mathpzc{w}_\pm^{(\gamma^\prime)}
\end{array}\right]\left[\begin{array}{c} f_\pm(\pmb{k},t)\\\\ g_\pm(\pmb{k},t)
\end{array}\right].
\end{equation}We stress that the  diagonal elements of  the  matrix in Eq.~(\ref{reduceddispersionequation}) are the dispersion relations given in Eq.~(\ref{disperisonlaws}). 

The solution of the above equation  can be written as a superposition of  eigenvectors of the linearized version of Eq.~(\ref{dispr1}). 
In fact, by introducing  the mixing angle   $\varphi\equiv\arctan\left(\frac{\chi}{1-\chi^2}\right)$ we  find
\begin{equation}
\left[\begin{array}{c}f_\pm(\pmb{k},t)\\ g_\pm(\pmb{k},t) \end{array}\right]=\frac{\mathpzc{C}_\pm^{(\gamma)}}{\sqrt{1+\tan^2(\varphi)}}\left[\begin{array}{c}1\\\tan(\varphi)\end{array}\right]e^{-i\mathpzc{w}_\pm^{(\gamma)}t}-\frac{\mathpzc{C}_\pm^{(\gamma^\prime)}}{\sqrt{1+\tan^2(\varphi)}}\left[\begin{array}{c}\tan(\varphi)\\ -1 \end{array}\right]e^{-i\mathpzc{w}_\pm^{(\gamma^\prime)} t}.
\end{equation}  The  constants  $\mathpzc{C}_\pm^{(\gamma,\gamma^\prime)}$  are  determined by  supposing an experimental setup which starts--at $t=0$--without a hidden-photon field but with 
an incoming electromagnetic probe beam of finite  amplitude   $f_{\pm}(\pmb{k},0)=[4\pi/(2\omega_{\pmb{k}})]^{\nicefrac{1}{2}}$. With this idea in mind,  
we  obtain  a system of algebraic equations for $\mathpzc{C}_\pm^{(\gamma,\gamma^\prime)}$, whose  solutions  allow us  to  express  the  flavor-like  
components as 
\begin{eqnarray}\label{afield}
f_\pm(\pmb{k},t)={\sqrt{\frac{4\pi}{2\omega_{\pmb{k}}}}\mathpzc{A}_\pm(\pmb{k},t)}e^{-i\omega_{\pmb{k}}t}\quad \mathrm{and}\quad g_\pm(\pmb{k},t)={\sqrt{\frac{4\pi}{2\omega_{\pmb{k}}}}\mathpzc{B}_\pm(\pmb{k},t)}e^{-i\omega_{\pmb{k}}t},
\end{eqnarray} where--in  the limit of  weak-mixing   $\chi\ll1$  and  $\varphi\simeq \chi$--the  wave amplitudes  $\mathpzc{A}_\pm(\pmb{k},t)$ and $\mathpzc{B}_\pm(\pmb{k},t)$ 
approach to the following expressions  
\begin{eqnarray}\label{resonantamplitudee}
\mathpzc{A}_\pm(\pmb{k},t)&\approx& \exp\left\{i(n_\pm-1)\omega_{\pmb{k}}t+i\chi^2\sin\left(\frac{n_\pm-1}{\chi^2}\omega_{\pmb{k}}t\right)\exp\left(-\frac{1}{\chi^2}\kappa_\pm t\right)\right.\nonumber\\&&\qquad\left.-\kappa_\pm t-\chi^2\left[1-\cos\left(\frac{n_\pm-1}{\chi^2}\omega_{\pmb{k}}t\right)\exp\left(-\frac{1}{\chi^2}\kappa_\pm t\right)\right] \right\},\\
\mathpzc{B}_\pm(\pmb{k},t)&\approx& \chi\left\{\exp\left(-\frac{1}{\chi}\kappa_\pm t\right)-\cos\left(\frac{n_\pm-1}{\chi^2}\omega_{\pmb{k}}t\right)+i\sin\left(\frac{n_\pm-1}{\chi^2}\omega_{\pmb{k}}t\right)\right\},
\end{eqnarray} with   $n_\pm$ and $\kappa_\pm$ given  in Eq.~(\ref{refractioninde}). 

The  modulus square of $\mathpzc{B}_\pm(\pmb{k},t)$ gives us  the conversion  probability, which turns out to be  intrinsically associated with the  exponentials 
responsible for the damping  of the corresponding electromagnetic waves due to  the photon-paraphoton  oscillations
\begin{equation}
\mathpzc{P}_{\gamma_\pm\to\gamma_\pm^\prime}(\tau)\simeq \chi^2\left\{1+\exp\left(-\frac{2}{\chi^2}\kappa_\pm \tau\right)-2\exp\left(-\frac{1}{\chi^2}\kappa_\pm \tau\right)\cos\left(\frac{n_\pm-1}{\chi^2}\omega_{\pmb{k}}\tau\right)\right\}\label{progene}.
\end{equation}Observe that this formula  has been  evaluated at  the pulse length $t=\tau$ of the external laser wave [Eq.~(\ref{externalF})]. Interestingly, it   
resembles the probability  of conversion resulting from a setup in which a  constant magnetic field drives the photon-paraphoton oscillations \cite{Ahlers:2007rd,Ahlers:2007qf}. 
Note that Eq.~(\ref{progene}) is characterized by  an oscillatory  pattern which tends to be exponentially suppressed as the pulse length $\tau$ of the laser  wave  is much larger 
than the characteristic time of the transition process  $\sim \chi^2 \kappa_\pm^{-1}$. In such a case,  Eq.~(\ref{progene}) asymptotically approaches $\mathpzc{P}_{\gamma_\pm\to\gamma_\pm^\prime}(\tau)\simeq\chi^2$. 
Conversely, when  the attenuation factors $\sim\kappa_\pm \tau/\chi^2$ and the trigonometric argument $(n_\pm-1)\omega_{\pmb{k}}\tau/\chi^2$ are much smaller than unity, the probability of conversion 
reduces to 
\begin{eqnarray}\label{ratefermionphotongeneration}
\mathpzc{P}_{\gamma_\pm\to\gamma_\pm^\prime}\approx\frac{1}{\chi^2}\left[(n_\pm-1)^2\omega_{\pmb{k}}^2+\kappa_\pm^2\right]\tau^2.
\end{eqnarray} 

It is worth mentioning that this expression coincides with the outcome resulting from perturbation theory when the  Abelian  fields in Eq.~(\ref{effectiveaction}) 
are canonically quantized.  In this context, the probability amplitude of the photon-paraphoton oscillation  can be read off directly from  the off-diagonal 
elements in Eq.~(\ref{eqIII}). Within the accuracy to the second order with respect to the radiative corrections, it is explicitly  given by 
\begin{equation}\label{generalamplitude}
T_{\mathpzc{e}^{(i)} k^\prime ,\mathpzc{e}^{(f)} k}=\frac{i}{\chi} \frac{\mathpzc{e}_\mu^{(f)}\Pi^{\mu\nu}(k,k^\prime)\mathpzc{e}_\nu^{(i)}}{2V(\omega_{\pmb{k}^\prime}\omega_{\pmb{k}})^{\nicefrac{1}{2}}}.  
\end{equation}Here  $V$ denotes  the normalization volume, whereas $\mathpzc{e}^{(i)}_\mu$ and $\mathpzc{e}^{(f)}_\mu$ are  the initial  and final polarization states,  respectively. We  
suppose  the former to be associated with  visible photons [Eq.~(\ref{chiralityexpansio})] so that $\mathpzc{e}^{(i)}=\Lambda_\pm$. In contrast, the respective   polarizations of the final hidden 
$\rm U(1)-$gauge states  are chosen as $\mathpzc{e}^{(f)}=\Lambda_\pm^*$. We insert these and the expression for the polarization tensor [Eq.~(\ref{ptcircular})] into the Eq.~(\ref{generalamplitude}). Consequently, 
the  probability amplitude becomes
\begin{equation}
T_{\gamma_\pm\to\gamma_\pm^\prime}=\frac{i}{\chi }\frac{\pi_\pm }{2V \omega_{\pmb{k}}}(2\pi)^4\delta^4(k^\prime-k).
\end{equation}Next, the modulus squared of this formula  is  integrated  over the  momentum of the  final hidden-photon field. Such a procedures allows us to write  
the conversion probability   as  in  Eq.~(\ref{ratefermionphotongeneration}), provided  the usual  interpretation of the interacting  time $\tau\equiv2\pi\delta(0)$ is used,  where the occurence of  
a vanishing argument in the Dirac delta  is a direct consequence  of the energy conservation in the  process. 

We want to conclude this section by deriving the potential observables in  an  optical  experiment  assisted by the   field 
of a circularly polarized plane wave.  To this end  we  emphasize  that,   in this  kind  of background,  the vacuum behaves as a chiral medium rather 
than an uniaxial material. In correspondence, the polarization  plane  of  an incoming linearly polarized probe beam  undergoes a  rotation  
$\vartheta(\tau)$ due to  the relative phase difference between the propagating (helicity) modes 
[Eq.~(\ref{afield})]:
\begin{eqnarray}\label{rotation}
&&\vert\vartheta(\tau)\vert \approx\frac{1}{2}\left\vert(n_+-n_-)\omega_{\pmb{k}}\tau+\chi^2\sin\left(\frac{n_+-1}{\chi^2}\omega_{\pmb{k}}\tau\right)\exp\left(-\frac{1}{\chi^2}\kappa_+ \tau\right)\right.\nonumber\\
&&\qquad\qquad\qquad\qquad\qquad\qquad-\left.\chi^2\sin\left(\frac{n_--1}{\chi^2}\omega_{\pmb{k}}\tau\right)\exp\left(-\frac{1}{\chi^2}\kappa_- \tau\right)\right\vert\ll1.
\end{eqnarray}As such, the  interaction with the strong field of the wave transforms the outgoing   probe beam into an elliptically polarized wave. In our context, the degree of 
ellipticity $\psi(\tau)$ is an outcome of   both the absorption of visible waves via the  production of pairs of  MCPs and their conversion into hidden-photons. Hence, it 
is  determined by the difference between the attenuation coefficients of the visible  circular modes [Eq.~(\ref{afield})]. Explicitly,
\begin{eqnarray}\label{ellipticity}
&&\vert\psi(\tau)\vert \approx\frac{1}{2}\left\vert(\kappa_--\kappa_+)\tau+\chi^2\cos\left(\frac{n_+-1}{\chi^2}\omega_{\pmb{k}}\tau\right)\exp\left(-\frac{1}{\chi^2}\kappa_+ \tau\right)\right.\nonumber\\
&&\qquad\qquad\qquad\qquad\qquad\qquad-\left.\chi^2\cos\left(\frac{n_--1}{\chi^2}\omega_{\pmb{k}}\tau\right)\exp\left(-\frac{1}{\chi^2}\kappa_- \tau\right)\right\vert\ll1.
\end{eqnarray} 

As in the case of the conversion probability, the oscillatory pattern in both observables is suppressed when  very long pulses  $\chi^2 \kappa_\pm^{-1}\ll\tau$ 
are considered. The resulting asymptotes coincide with the  standard results for the ellipticity and rotation in a pure MCPs model  
\cite{Villalba-Chavez:2013gma}. However, the effects resulting from the photon-paraphoton oscillations   could be quite noticeable if the pulse length 
is much smaller than  $\chi^2 \kappa_\pm^{-1}$ and if it  simultaneously  satisfies  the condition $\tau\ll\chi^2 \omega_{\pmb{k}}^{-1}(n_\pm-1)^{-1}$. Then 
Eqs.~(\ref{rotation}) and (\ref{ellipticity}) approach
\begin{eqnarray}\label{approachobservables}
&&\vert\vartheta(\tau)\vert \approx\left\vert(n_+-n_-)\omega_{\pmb{k}}\tau+\frac{1}{4\chi^2} \left[(n_--1)\kappa_--(n_+-1)\kappa_+\right]\omega_{\pmb{k}}\tau^2\right\vert,\\
\vert\psi(\tau)\vert& \approx &\left\vert(\kappa_--\kappa_+)\tau+\frac{1}{4\chi^2}\left[(n_--1)^2-(n_+-1)^2\right]\omega_{\pmb{k}}^2\tau^2+\frac{1}{4\chi^2}\left(\kappa_+^2-\kappa_-^2\right)\tau^2 \right\vert.
\end{eqnarray} Interestingly, the leading terms  are increased by a  factor two as compared to those  associated with the pure MCPs model. Such a feature
provides an  evidence that, in the limit under consideration, the production of pairs of MCPs and the vacuum birefringence are  
stimulated by the existence of a hidden-photon field.

\section{Experimental prospects   \label{PES}}

\subsection{Estimating the exclusion limits   \label{PAPH}}

Restrictions on the $(\epsilon,m_\epsilon)$  plane can be established whenever in certain confidence levels $\psi_{\mathrm{CL}\%},\  \vartheta_{\mathrm{CL}\%}$, 
neither rotation of the polarization plane [Eqs.~(\ref{rotation})] nor  ellipticity of the outgoing probe beam  [Eq.~(\ref{ellipticity})]  are  detected.  
We note that, while experimental data for the proposed setup  do not exist yet, ellipticities and rotation angles can nowadays be  measured with an accuracy 
of about $\sim10^{-10}$\,rad in the optical regime \cite{Muroo_2003}.\footnote{An experiment to measure  vacuum birefringence by probing a Petawatt optical laser [$\xi\gg1$] 
with a x-ray free electron laser,  has been proposed by the HIBEF consortium \cite{HIBEF}.} Hereafter, we  give first estimates of the exclusion bounds resulting   from  the absence 
of the aforementioned signals  by taking the previous value as  reference for the sensitivity parameters $\psi_{\mathrm{CL}\%},\  \vartheta_{\mathrm{CL}\%}$.  
Accordingly, we have to  solve the inequalities  $\psi_{\rm CL\% }>\psi(\tau)$ and $\vartheta_{\rm CL\% }>\vartheta(\tau)$  but, due to the very complicated 
dependence of $\psi(\tau)$ and $\vartheta(\tau)$  on the unknown parameters of our  theory  $\epsilon,\ m_\epsilon\ \mathrm{and}\ \chi$,  no analytic solutions can 
be derived. Therefore, we rather   determine  their  bounds  numerically. However, in  doing so we should keep in mind that the  application  of the expressions obtained 
so far  requires  an external laser field  which  approaches to   our monochromatic  plane-wave model [Eq.~(\ref{externalF})]. In practice,  the monochromaticity  of the  
high-intensity laser wave can be implemented as long as the laser-source emits a pulse  with an  oscillation period $\mathpzc{T}=2\pi\varkappa_0^{-1}$  much 
smaller than its temporal length $\tau$, i. e.,  $\varkappa_0\tau\gg1$. Regarding the plane-wave character, it formally implies that the long-laser wave is infinitely 
extended in the plane perpendicular to the propagation direction. However, in an actual experimental realization,  this condition can be  considered as  satisfied when the waist size 
of the laser beam $w_0$ is much greater than its wavelength [$w_0\gg\lambda_0$ with $\lambda_0=2\pi\varkappa_0^{-1}$]. 

In order to satisfy both conditions,  we choose, for our external laser field, the set of parameters associated with the Petawatt High-Energy Laser for heavy Ion eXperiments 
(PHELIX) \cite{PHELIX}, currently  under  operation at GSI in Darmstadt, Germany. We are particularly interested in the nanosecond frontend of  PHELIX [$w_0\approx 100-150 \mu \mathrm{m}$],  
since  it   operates with an infrared wavelength $\lambda_0\simeq1053\ \rm nm$ [$\varkappa_0\simeq1.17\ \rm eV$]  and can reach a peak intensity $I\simeq 10^{16}\  \rm W/cm^2$, 
corresponding to   a  parameter $\xi\simeq 6.4\times 10^{-2}$ in a pulse length $\tau\simeq20 \ \rm ns$.  In addition, we  will  study   the results coming from the  specification 
of the long  high-energy pulse of  $400\ \rm J$   at the Laboratoire pour l'Utilisation des Lasers Intenses (LULI) \cite{LULI}--currently in operation at Palaiseau, France. Similarly 
to the previous external source,  we will focus ourselves on  the  nanosecond facility at  LULI(2000) [$w_0\sim100 \ \mu\mathrm{m}$], which can operate with  the same  central 
frequency as PHELIX  once its fundamental harmonic is used. However, its  pulse length is shorter  $\tau\simeq 1.5-4 \ \rm ns$  and its  maximum  intensity is  lower  
$I\simeq6\times 10^{14}\ \rm W/cm^2$ [$\xi\simeq2\times 10^{-2}$].  Regarding the probe beam, we suppose that it is an optical laser obtained by coupling out a tiny fraction of the external 
laser wave whose intensity turns out to be much weaker than the intensity of the strong beam.  We  assume, in particular, that  the probe  
frequency can be  doubled  [$\omega_{\pmb{k}}=2\varkappa_0=2.34\ \rm eV$] afterwards. 

\begin{figure}
\includegraphics[width=.48\textwidth]{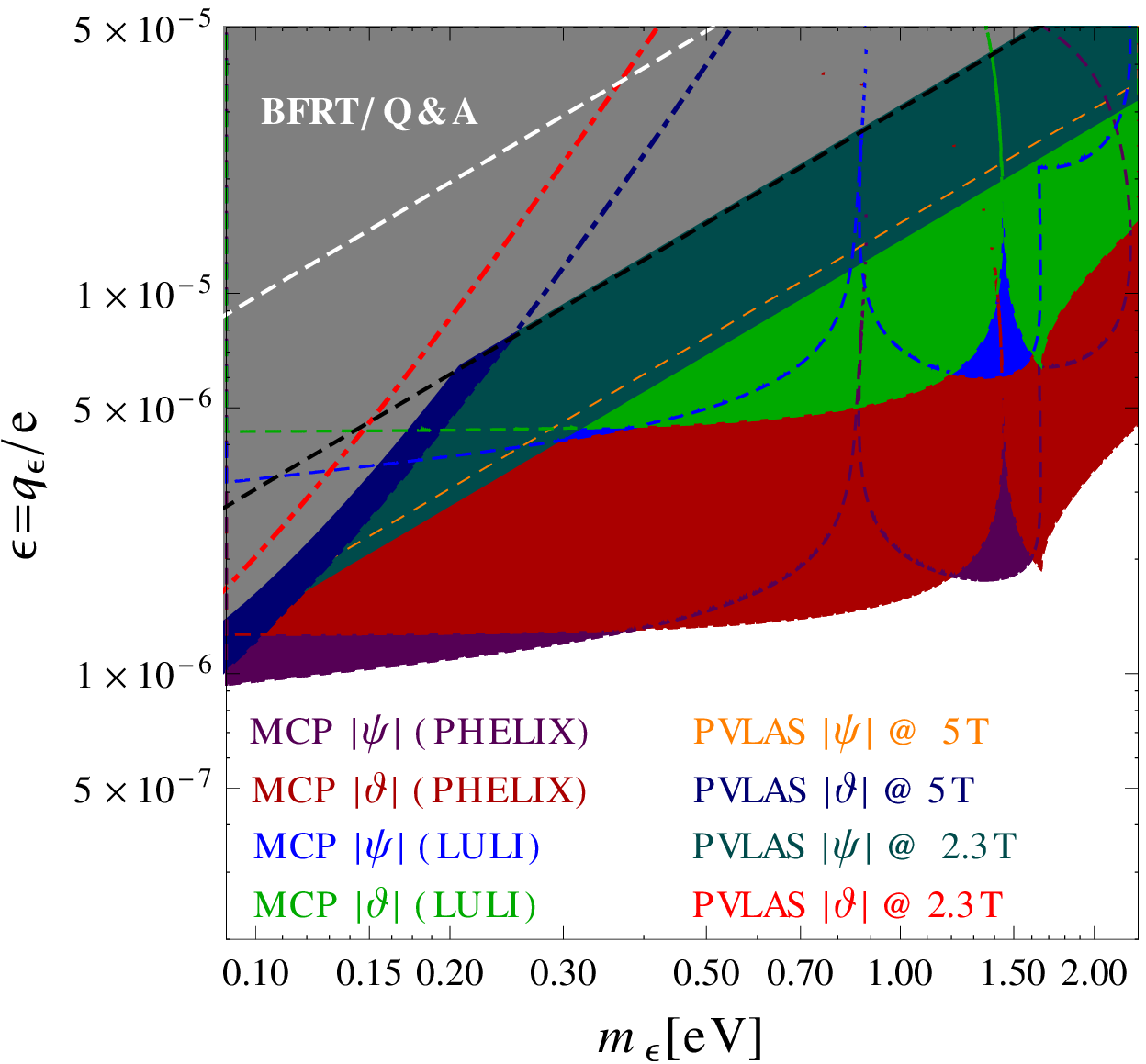}
\includegraphics[width=.49\textwidth]{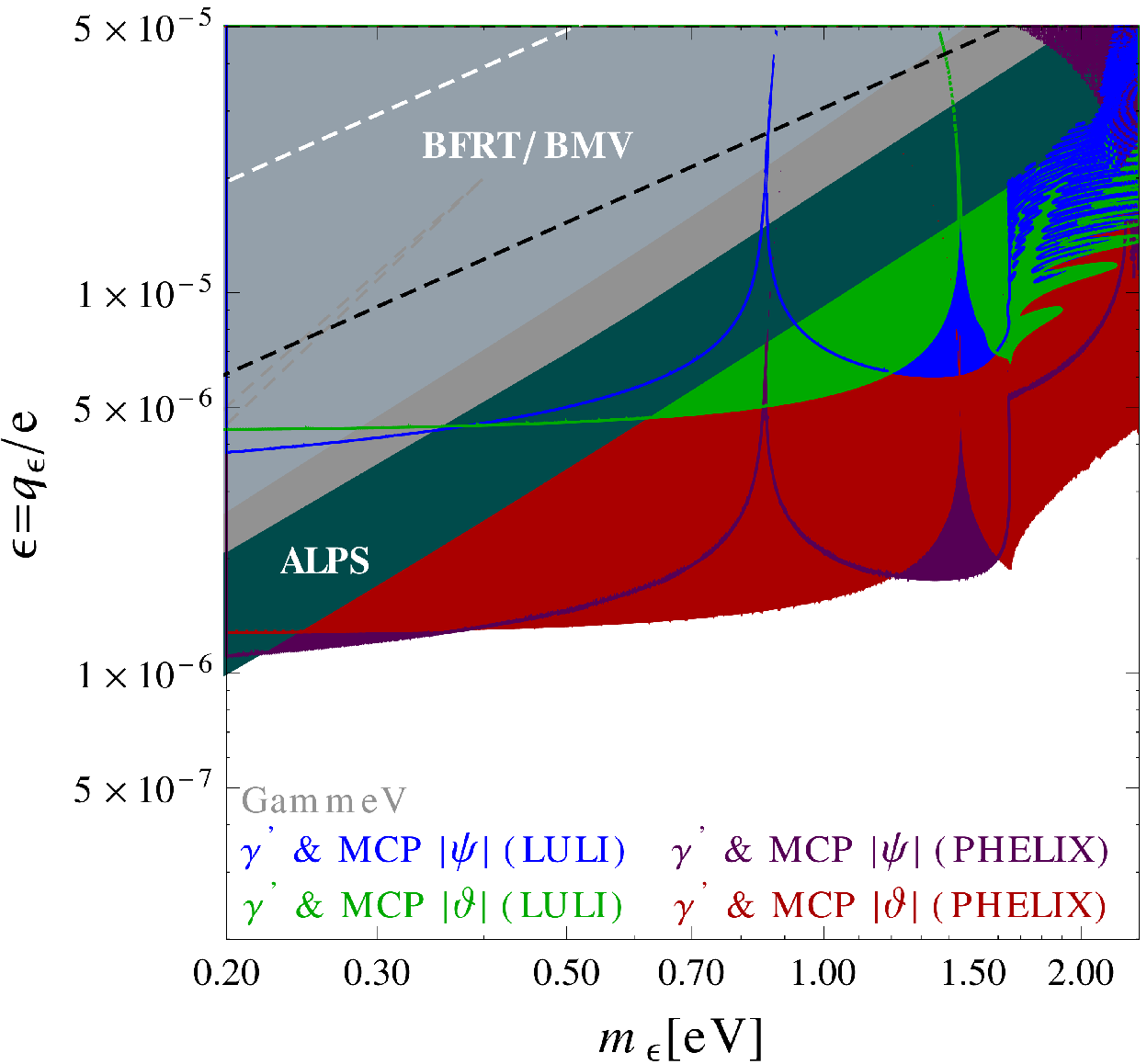}
\caption{\label{fig:mb003}Estimates of  constraints for MCPs  of  mass $m_\epsilon$ and relative coupling constant $\epsilon$ derived  from the absence of signals in  
a plausible polarimetric setup assisted by a circularly polarized  laser field of moderate intensity. While the left panel provides the results associated with MCPs, 
the right one shows the outcomes of the  model including  a hidden-photon field $(\gamma^\prime)$.  In both panels the white (LULI) and black (PHELIX) dashed lines correspond to the expression 
$\xi_\epsilon=\epsilon m \xi/m_\epsilon=1$.  The picture in the left, includes the  exclusion regions  coming  from  various  experimental  collaborations searching 
for rotation and ellipticity in a constant magnetic field  such as BFRT \cite{Cameron:1993mr},  PVLAS \cite{Zavattini:2007ee,DellaValle:2013xs}  and Q $\&$ A \cite{Chen:2006cd}.  
However, the  shaded areas in the upper left corner in the right panel result from   different experimental collaborations dealing with the Light Shining Through a Wall 
mechanism. The respective $95\%$  confidence  levels needed to reproduce these results are summarized  in Refs.~\cite{Ahlers:2007qf,Ehret:2010mh}. }
\end{figure}

Observe that,  with the above  assumptions, the photo-production of an electron-positron pair  cannot take  place. In the fields  under consideration, the occurrence of a linear 
Breit-Wheeler reaction would require  probe photons with energy greater than the threshold value $\omega_{\pmb{k}}\gtrsim m^2\varkappa_0^{-1}\approx10\ \rm GeV$. As a consequence, 
no contribution other than the one  induced by the decay of the probe beam into MCPs pairs and its oscillation into a  paraphoton  is expected in the dichroic effect [Eq.~(\ref{ellipticity})]. 
Therefore, an eventual  detection  of  ellipticity  in the outgoing  probe beam  can be  understood as  a clear manifestation of  physics beyond SM.   Furthermore,  below  the first
pair production   threshold [$k\varkappa\ll2m^2$] and for $\xi<1$,  the  birefringence  of the  pure QED vacuum  is predicted to be extremely weak \cite{baier,Villalba-Chavez:2013gma}. 
Indeed, the  rotation of the polarization plane  for  spinor QED is given by 
\begin{equation}
 \vert\vartheta_{\rm QED}\vert\approx\frac{2}{315}\frac{\alpha}{\pi}\frac{(\varkappa k)^3}{m^4\omega_{\pmb{k}}}\xi^2 \tau.
\end{equation}When  evaluating this expression with the parameters of the nanosecond frontend of  PHELIX in a counter propagating geometry [$\pmb{k}\parallel-\pmb{\varkappa}$],  we find that its contribution 
 $\vert\vartheta_{\rm QED}\vert\sim10^{-21}\ \mathrm{rad}$ turns out to be extremely tiny in comparison with  the reference 
 sensitivity $\sim 10^{-10}\ \rm rad$ to be used henceforth.  That the pure QED effect is so tiny  is important for practical purposes, because it would  allow for 
 isolating polarization-dependent  effects stemming only from the self-interaction of the electromagnetic 
field in vacuum mediated by  hypothetical degrees of freedom. Therefore, also   by  sensing a rotation  in the polarization plane [Eq.~(\ref{rotation})]  
other than the predictions  coming from  QED,  we  can   probe the existences of our dark matter candidates.

The  first estimated  of the described  settings are shown in  Fig.~\ref{fig:mb003} for the  particular situation in which the collision is head-on, i. e., 
$\pmb{k}\cdot\pmb{\varkappa}=-\omega_{\pmb{k}}\varkappa_0$.  Our exclusion regions are shaded in purple and red for PHELIX and in blue and green  for LULI. 
We emphasize that they are expected to be trustworthy when the bounds are below  the respective  dashed line--white for LULI and black for PHELIX--corresponding to  $\xi_\epsilon=\epsilon m \xi/m_\epsilon=1$.  
While the left panel  displays the results coming from the pure MCP model, the discovery potential including the paraphoton effects is shown in the right panel. 
Note that, in the latter, the shaded regions   were  derived by considering the relation $\chi=\epsilon$, so that a direct  comparison with the pure MCP model 
can be established.  In contrast to the left panel,  the one in the  right   does not show the constraints coming  from the experimental searches of polarimetric 
signals. Instead, we have incorporated the upper bound obtained from the ALPS collaboration \cite{Ehret:2010mh} which--at the time of writing--turns out to be the 
most stringent laboratory-based limit for MCPs in a model with massless paraphotons. Also, included are the exclusion regions resulting  from other experimental 
collaborations  dealing with the Light Shining Through a Wall mechanism such as BMV \cite{Robilliard:2007bq}, BFRT \cite{Cameron:1993mr} and GammeV \cite{Chou:2007zzc}. 
Observe that  in the vicinity of the first threshold mass  is $m_1\approx1.64\ \rm eV$, \footnote{Note that for the laser parameters used here, the second contribution 
in the threshold mass [Eq.~(\ref{thresholdmass})], i. e.  $\epsilon^2m^2\xi^2\sim  10^{-3}\ \rm eV^2$ is neglectable in comparison with the first term $k\varkappa/2$.}  the upper limits  resulting  from the search of the rotation angle are  more 
restrictive than those arising  from the ellipticity.  Nearby, the bounds   are $\epsilon< 1.9 \times 10^{-6}$ for PHELIX  and $\epsilon< 6.5\times 10^{-6}$ for LULI. 
Also, Fig.~\ref{fig:mb003}  verifies the statement given below Eq.~(\ref{thresholdmass}) about the contribution of higher thresholds, since the picture covers a region 
including  the second threshold mass $m_2\approx2.34\ \rm eV$,  at which the  upper limit  undergoes  a  relaxation. 

\begin{figure}
\includegraphics[width=6.3 in]{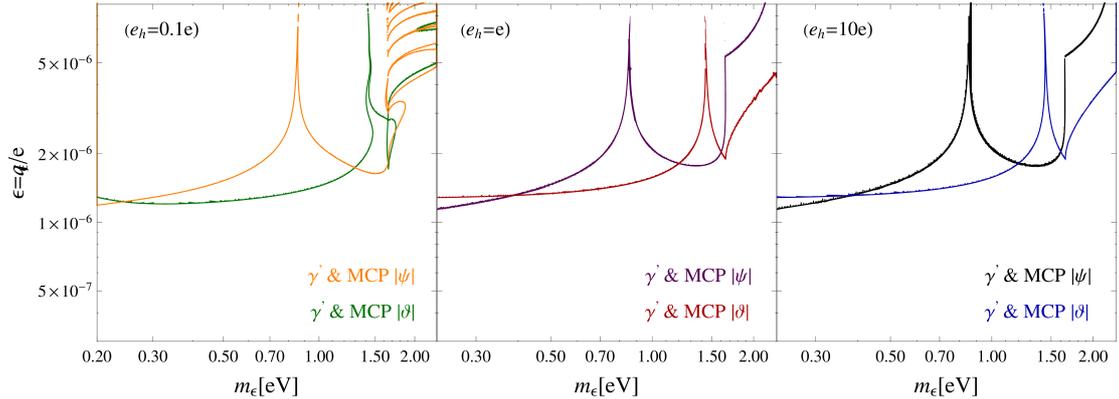}
\caption{\label{fig:mb004}Parameter space to be  ruled out for MCPs in a model with paraphotons $(\gamma^\prime)$. The  expected exclusion limits have been 
obtained by assuming the  absence of signals in a  polarimetric setup assisted by a circularly polarized wave associated with the nanosecond frontend of the  PHELIX 
laser. Here  the constraints on the kinetic mixing parameter for various values  of the hidden coupling constant are displayed by contour lines [see legend]. }
\end{figure}

In the right panel of this figure, we see  that the constraints  coming  from the  plausible absence of  signals in our  LULI setup  follow a path very similar to the one obtained from the 
pure MCPs model for  masses below the first threshold mass $m_1\approx1.64\ \rm eV$. This fact manifests a dominance of the first contributions to the observables [Eqs.~(\ref{rotation}) 
and (\ref{ellipticity})] due to a plausible exponential suppressions of the paraphoton terms. Hence, we deduce that  in such a region,  the  characteristic times involved in the respective 
damping factors of the waves $\chi^2\kappa_{\pm,1}^{-1}$  turn out to be much smaller than the laser pulse lengths $\tau\gg \chi^2\kappa_{\pm,1}^{-1}$. However, for masses 
embedded in the range $m_1<m_\epsilon<m_2$, the upper bounds  resulting  from  LULI's parameters   are characterized by fluctuating patterns which are   absent in a pure MCPs scenario. 
The occurence of these trends  is closely associated with the photon-paraphoton  oscillations.  In contrast to  masses below $m_1$, the region in which $m_1<m_\epsilon<m_2$ 
turns out to feature characteristic times $\chi^2\kappa_\pm^{-1}$  much  larger  than (in the case of LULI)--or at least of  the order  of (in the case of PHELIX)--the  pulse lengths $\tau$ used. 
This is  caused  by the contributions coming from the second threshold [Eq.~(\ref{absorptioncoefficient2order})] which become--in the region under consideration--the 
leading order term in the absorption coefficients. 

Actually, the dependence of the hidden gauge coupling  $e_h$ introduces a certain level of uncertainty.  Fig.~\ref{fig:mb004}  shows how the constraints for PHELIX might vary  
as  $e_h$  changes by  an order of magnitude  around the natural value  $\vert e_h\vert =\vert e\vert$. Observe that, these variations are almost imperceptible for masses below the first threshold mass.  
However, slight  deviations  in  the  exclusion  bounds  are  displayed in a vicinity of the first threshold and within the interval  where the photon-hidden-photon oscillations 
are more pronounced. Clearly,  these  first estimates  indicate that experiments driven by long laser pulses of moderate intensities might be sensitive in regions of the parameter 
space which are not  excluded  by the outcomes of current  laboratory-based collaborations such as PVLAS and BFRT. Particularly, in a vicinity of the  
first threshold mass,  the present  upper bound might be an order of magnitude more stringent than the one resulting from the PVLAS and ALPS analyses.

\subsection{Identification of promising scenarios    \label{DSM}}

So far we have investigated the plausible situation in which no  optical  change  is detected. In this subection, we  shall examine the  
case  where  the  induced  ellipticity and rotation of the outgoing  probe beam--due to the  vacuum polarization effects of  MCPs and 
a hidden-photon field--become manifest.  In first instance,  a measurement of the absolute value  of the aforementioned observable should 
be enough  for finding the values  of $\epsilon$ and $m_\epsilon$ provided that only  pure MCPs are realized in nature. However, with 
the inclusion of the hidden-photon field the mixing parameter $\chi$ emerges and consequently, the polarimetric measurements by themself  
do not unambiguously determine the unknown  particle attributes. This situation might even become worse if other dark matter candidates 
such as axion-like particles would exist as well  at the energy scale relevant for MCPs and paraphotons.  Valuable information is  however at our disposal:  
by  investigating the signal dependencies on the available experimental  quantities such as  the  intense laser parameter $\xi$, the temporal 
length $\tau$ and the  wave length of the probe beam $\lambda=2\pi\omega^{-1}$ one can  establish the phenomenological differences that 
result from the models under consideration. 

\begin{figure}
\includegraphics[width=6 in]{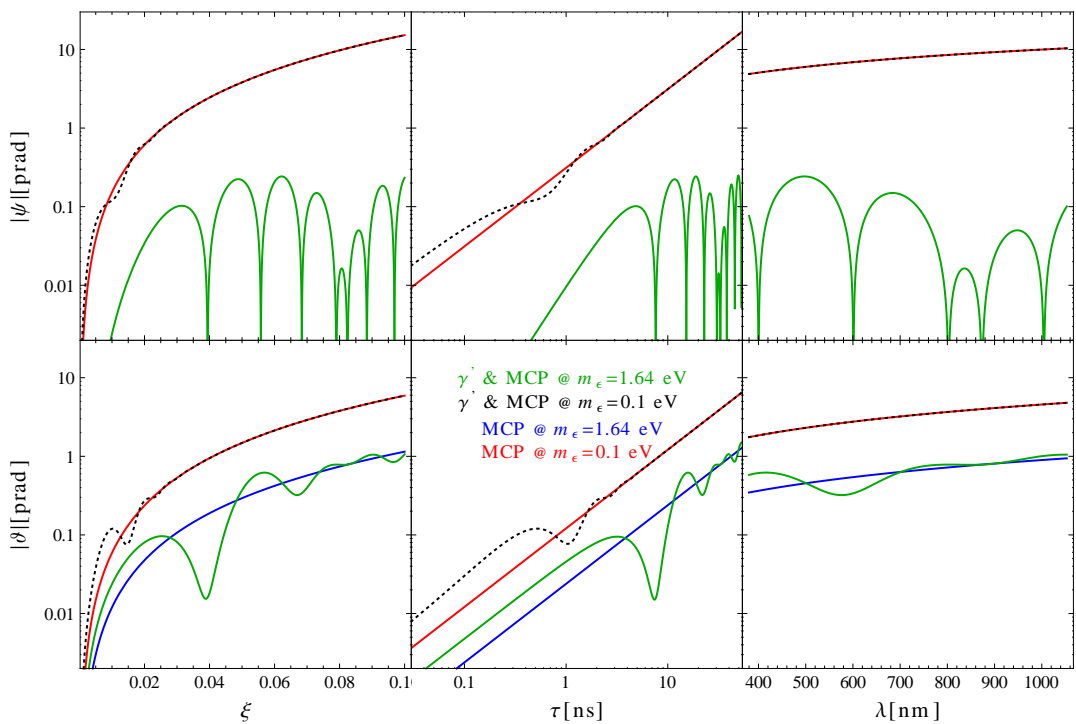}
\caption{\label{fig:mb005} Dependence of the absolute value of the ellipticity $\vert\psi\vert$ [upper panels] and rotation angle $\vert\vartheta\vert$  
[lower panels] on the  intensity 
parameter $\xi$ [left panel], pulse length $\tau$ [central panel] and wavelength of the probe $\lambda$ [right panel]. As a benchmark point we assume a massless hidden 
photon field with kinetic mixing parameter $\chi=5\times 10^{-7}$ and hidden coupling $e_h=e$.  In each plot the remaining external  parameters are 
kept at $\xi=6.4\times 10^{-2}$, $\tau=20\ \rm ns$, $\pmb{k}\parallel-\pmb{\varkappa}$, $\lambda=\lambda_0/2$ with $\lambda_0=2\pi\varkappa_0^{-1}=1053\ \rm nm$  the wavelength 
of the intense laser field.  Here  the outcomes  resulting  from a  pure MCP model at  $m_\epsilon=0.1\ \rm eV$ are shown in red, whereas  the respective 
patterns at the first threshold mass $m_1\approx(k\varkappa/2)^{\nicefrac{1}{2}}$ are  in blue. The curves in green  and  dotted black  were obtained by   
including the paraphoton field. They also correspond to the case in which the  mass of the minicharges are $m_\epsilon=m_1$ and  $m_\epsilon=0.1 \ \rm eV$, respectively.}
\end{figure}

Fig.~\ref{fig:mb005} summarizes  the  behavior of the signals not only when the MCP  mass coincides with the first  threshold mass $m_\epsilon=m_1\approx(k\varkappa/2)^{\nicefrac{1}{2}}$ but 
also at  $m_\epsilon=0.1\ \rm eV$. The  corresponding results associated with the pure MCP scenario are plotted in blue and red, whereas the outcomes  
including the effects of a hidden-photon field are shown in green and black dotted curves. All these results were derived by using the  benchmark 
parameters of the nanosecond frontend of PHELIX  [$\xi=6.4\times 10^{-2}$, $\tau=20\ \rm ns$, $\lambda_0=1053\ \rm nm$] and by considering a 
probe beam  with $\lambda=\lambda_0/2$  colliding  head on with the intense laser wave.  Generally speaking we find that  both signals tend 
to grow with the increase  of the external laser attributes.  At $m_\epsilon=0.1\ \rm eV$  and small values of  the intensity parameter $\xi<4\times10^{-2}$, 
the dependencies of the  ellipticity $\vert\psi\vert$ [upper panels] and rotation angle $\vert\vartheta\vert$ [lower panels]   show   slight differences  
between the pure MCP model and the scenario dealing with the paraphoton effects.  Precisely in this region  the characteristic time of the transition 
process  $\sim \chi^2 \kappa_\pm^{-1}$ becomes much bigger than the pulse length [$\tau=20\ \rm ns$]  and the oscillatory patterns due to the photon-paraphoton 
oscillation alters the signal compared to a pure MCPs model. Conversely, for values of  $\xi>4\times10^{-2}$,  $\chi^2 \kappa_\pm^{-1}$ turns out to 
be smaller than $\tau$ leading to  exponential suppressions of the hidden-photon effects described in Sec.~\ref{CPPO}. A similar behavior occurs for 
fixed  $\xi=6.4\times 10^{-2}$ as  the pulse length  $\tau$  varies. Indeed, the central panel in Fig.~\ref{fig:mb005} shows that at $m_\epsilon=0.1\ \rm eV$ 
and duration smaller than  $\lesssim 6\ \rm ns$ the signal starts to be sensitive for hidden-photon  effects since the characteristic time turns out to be 
greater than  the pulse length of the strong laser wave.  Observe that,  at $m_\epsilon=0.1 \ \rm eV$, the respective dependencies of the observables on 
the wavelength of the probe beam do not reveal any differences. This is because, for the remaining benchmark parameters, the pure MCP contributions turn 
out to be  dominant.

\begin{figure}
\includegraphics[width=6 in]{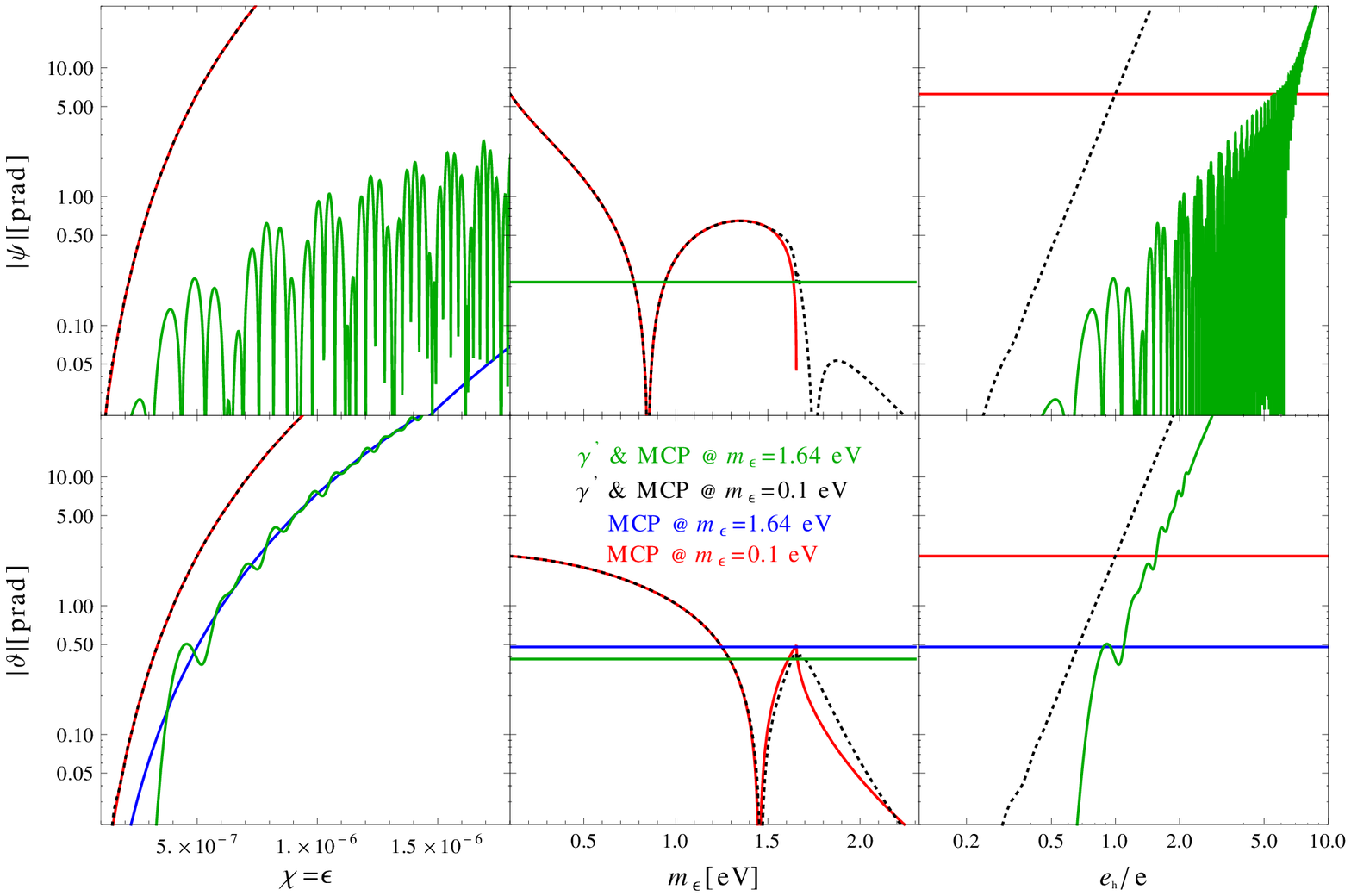}
\caption{\label{fig:mb006} Dependence of the absolute value of the ellipticity $\vert\psi\vert$ [upper panels] and rotation angle $\vert\vartheta\vert$  
[lower panels] on the  kinetic mixing parameter $\chi$ [left panel], mass $m_\epsilon$ [central panel] and the relative hidden coupling $e_h/e$ [right panel]. 
The same  benchmark values of  Fig.~\ref{fig:mb005}--$\xi=6.4\times 10^{-2}$, $\tau=20\ \rm ns$, $\pmb{k}\parallel-\pmb{\varkappa}$, $\lambda=\lambda_0/2$,  
$\lambda_0=2\pi\varkappa_0^{-1}=1053\ \rm nm$--have been used.}
\end{figure}

The signal  drastically changes at the first threshold mass.  Here the minicharges tend to be  produced at rest [$\mathpzc{v}_1\to 0$], and the leading 
order terms of the absorption coefficients [Eqs.~(\ref{fermionpositivekappa}) and (\ref{fermionnegativekappa})]  decrease as   $\kappa_{+,1}\propto \mathpzc{v}_1$ and   $\kappa_{-,1}\propto \mathpzc{o}(\mathpzc{v}_1^2)$, respectively  
\cite{Villalba-Chavez:2013gma}. Certainly, at this point,  the contributions coming from  the second threshold, i. e. $\kappa_{\pm,2}\propto\xi_\epsilon^4$ 
[see Eq.~(\ref{absorptioncoefficient2order})]  might also be important. Whatever be the dominant case, the main outcome would be a noticeable increment 
in the characteristic times  $\chi^2\kappa_\pm^{-1}$,   which can reach values  much larger than the corresponding pulse length $\tau$. In such a situation,  
the damping factors in Eq~(\ref{ellipticity}) can be approached by unity, the term  associated with the pure MCP model becomes negligible\footnote{\label{why3} 
This is  why the blue curve does not appear in the upper  panels associated with the ellipticity.} and the ellipticity  follows a fluctuating pattern
\begin{eqnarray}\label{ellipticityosc}
&&\vert\psi(\tau)\vert \approx\frac{1}{2}\chi^2\left\vert \cos\left(\frac{n_+-1}{\chi^2}\omega_{\pmb{k}}\tau\right)-\cos\left(\frac{n_--1}{\chi^2}\omega_{\pmb{k}}\tau\right)\right\vert.
\end{eqnarray}The outcomes displayed in the upper panel of Fig.~\ref{fig:mb005} clearly highlight this trend. According to Eq.~(\ref{ellipticityosc}), such a pattern is a 
direct consequence of the photon-paraphoton transitions, whose oscillation probabilities [Eq.~(\ref{progene})] reduce  to the expressions  
$\mathpzc{P}_{\gamma_\pm\to\gamma_\pm^\prime}(\tau)\approx  4\chi^2\sin^2\left(\frac{\mu_\pm^2}{4 \omega_{\pmb{k}}}\tau\right)$ with $\mu_\pm^2=4\frac{n_\pm-1}{\chi^2}\omega_{\pmb{k}}^2$ 
which resemble the one resulting from the  massive paraphoton theory \cite{Ahlers:2007rd,Ahlers:2007qf}.  Regarding the behavior of the  rotation angle [lower panel in Fig.~\ref{fig:mb005}],  the  
situation is slightly different. Based on similar arguments,  we found that  Eq~(\ref{rotation}) approaches to  
\begin{equation}\label{rotaosc}
\vert\vartheta(\tau)\vert \approx\frac{1}{2}\left\vert(n_+-n_-)\omega_{\pmb{k}}\tau+\chi^2\left[\sin\left(\frac{n_+-1}{\chi^2}\omega_{\pmb{k}}\tau\right)-\sin\left(\frac{n_--1}{\chi^2}\omega_{\pmb{k}}\tau\right)\right]\right\vert.
\end{equation}Here the occurrence of  fluctuations is also assignable to the photon-paraphoton oscillations. However,   in contrast to our previous analysis,  
the  standard result  for a model without  paraphoton -- first term in the above equation -- remains  important  and even dominant as the external parameters 
increase. 

In Fig.~\ref{fig:mb006}, we plot the ellipticity and rotation of the polarization plane with respect to the   unknown parameters of the theory. Particularly, the left and right  panels reveal 
how the signals might change  with the  mixing parameter $\chi$ and the relative hidden coupling $e_h/e$. From the former we  note  that at  $m_\epsilon=0.1 \ \rm eV$ 
the theory including a hidden-photon field follows the path dictated  by the  pure MCPs model.  Besides, a  fast decrease in both signals can be observed for small 
values of $\chi$. This trend is also manifest with respect to $e_h/e$  at the same reference mass [black dotted curve]. Here,  the outcome resulting 
from  the pure MCP scenario [horizontal red and blue lines] are not sensitive to  variations  of the relative hidden coupling because the latter  only arises within  the framework 
of a hidden-photon model.  In both--left and right--panels, one  recognizes the fluctuating  patterns [Eqs.~(\ref{ellipticityosc}) and (\ref{rotaosc})]  induced by the 
photon-paraphoton oscillations  at the first threshold mass $m_1=1.64\ \rm eV$. It can be seen  that the  oscillations caused by the photonaraphoton coupling  tend to be less pronounced 
as $e_h/e$ increases. The central panel of this figure  displays how both observables depend on the mass $m_\epsilon$ of MCPs.  There, in blue and green lines  are  indicated the reference values  
obtained for a fixed mass $m_\epsilon=m_1$.  The ellipticity 
resulting from  this  scenario  clearly shows the discontinuity at the first threshold mass discussed in Sec.~\ref{spectraldecompos} and associated with the nature of the 
absorption coefficients. This  feature is smoothed when a hidden photon field is considered [see dotted black curve].  At the first threshold, this observable  is  constant 
in  both scenarios, with the particularity of being extremely tiny  in the pure MCPs model [see footnote~\ref{why3}]. Conversely, the dependence of $\vert\vartheta(\tau)\vert$ 
with respect to the mass  $m_\epsilon$ follows a continuous path in both contexts and only slight differences appear for masses  above the first threshold mass $m_1$. This last 
observation  could be anticipated based on the analyses  of  our  previous discussions.

\section{Conclusions}

Accurate polarimetric techniques, searching for the birefringence and dichroism of the quantum  vacuum polarized by the field of a laser pulse, can be powerful probes for  
testing some effective theories beyond SM. We have considered the particular situation in which the external laser wave is circularly polarized and extended the 
results derived in Ref.~\cite{Villalba-Chavez:2013gma} by incorporating  the effects induced by paraphotons. In order to polarimetrically verify the realization 
of the considered models it is essential to gain  detailed  information on their respective   phenomenologies. As such, one of the main  goals of this work was to 
provide features  which can allow us to distinguish  between the pure MCPs scenario and the coexistence of   a hidden-photon field. Throughout, 
we noted that  the  possibility of exchanging  photons with the external wave renders  the description of the problem  more cumbersome than in the case  of a static 
magnetic field. These nontrivial properties, in conjunction  with  the energy-momentum balance lead to the appearance of thresholds closely  associated with  a  
hypothetical  photo-production of pairs of MCPs, their  masses being determined by the frequencies of both laser fields. For $\xi_\epsilon<1$ and near  the first 
threshold, the chiral activity of the ``medium''  turns out to be quite pronounced and the searches of very light  MCPs and hidden-photons by using polarimetric 
setups  appear promising.  In connection, the induced ellipticity and rotation of the  polarization plane  of the probe beam were determined. 
When evaluating such observables  with the attributes of  modern  laser systems,  stringent constraints on the parameter spaces  were  found under the assumption 
of no signal detection.  These first  estimates  reveal  that a laser wave with a long temporal length and a moderate intensity might be an external-field source  suitable 
for searching very light weakly interacting particles with masses in the eV range. As such our  outcomes agree with and complement the  results obtained  in a previous  investigation  developed  
within the context of axion-like particles \cite{Villalba-Chavez:2013goa}.


\end{document}